\documentclass{article}
\usepackage[english]{babel}

\usepackage[letterpaper,top=2cm,bottom=2cm,left=3cm,right=3cm,marginparwidth=1.75cm]{geometry}

\usepackage{amsmath}
\usepackage{graphicx}
\usepackage[colorlinks=true, allcolors=blue]{hyperref}

\usepackage{hyperref}

\usepackage[round,sort]{natbib}
\usepackage{bm}
\usepackage[noend]{algpseudocode}
\usepackage{algorithmicx,algorithm}
\usepackage{booktabs}
\usepackage{multirow}
\usepackage{threeparttable}
\usepackage{bigstrut}
\date{}
\title{\textbf{Online Updating Huber Robust Regression for Big Data Streams}}

\author{Chunbai Tao and Shanshan Wang\\School of Economics and Managment, Beihang University}

\begin{document}
\maketitle

\begin{abstract}
Big data streams are grasping increasing attention with the development of modern science and information technology. Due to the incompatibility of limited computer memory to high volume of streaming data, real-time methods without historical data storage is worth investigating. Moreover, outliers may occur with high velocity data streams generating, calling for more robust analysis. Motivated by these concerns, a novel Online Updating Huber Robust Regression algorithm is proposed in this paper. By extracting key features of new data subsets, it obtains a computational efficient online updating estimator without historical data storage. Meanwhile, by integrating Huber regression into the framework, the estimator is robust to contaminated data streams, such as heavy-tailed or heterogeneous distributed ones as well as cases with outliers. Moreover, the proposed online updating estimator is asymptotically equivalent to Oracle estimator obtained by the entire data and has a lower computation complexity. Extensive numerical simulations and a real data analysis are also conducted to evaluate the estimation and calculation efficiency of the proposed method.

\end{abstract}

\providecommand{\keywords}[1]
{
  \textbf{\text{Keywords: }} #1
}

\\

\section{Introduction}

\subsection{Background}

In recent years, big data has become a significant area of research and garnered extensive attention across various industries, including finance, manufacturing, and astronomy. Compared with traditional datasets, big data mainly has “5V” characteristics, i.e., high Volume, high Velocity, high Variety, low Veracity, and high Value\citep{jinSignificanceChallengesBig2015}. For big data streams, namely big data generated sequentially in a stream style, the characteristic of  high Volume and high Velocity are particularly prominent, posing several challenges in its analysis. One such challenge concerns computer memory limitations. As big data streams are generated with such big volume, it is unrealistic to be completely stored in computer memory. Consequently, standard statistical analysis methods that assume adequate computer capacity are no longer applicable in the current settings. The second challenge pertains to the presence of outliers in big data streams, which can arise due to recording errors in the fast data-generating process. While linear regression-based methods are often used for big data analysis, they are highly sensitive to outliers, thereby limiting their effectiveness in this context. As such, it is imperative to develop computational efficient, memory-conserving and robust methods for big data stream analysis that can address these challenges.

\subsection{Related Works}

To address the storage limitation problem in big data analysis, many scholars have designed different methodologies , which can be loosely divided into three categories, i.e. subsampling-based approaches, divide-and-conquer approaches and online updating approaches\citep{wangStatisticalMethodsComputing2016}. According to \citeauthor{wangStatisticalMethodsComputing2016}(\citeyear{wangStatisticalMethodsComputing2016}), classic subsampling methods include bags of little bootstrap\citep{kleinerScalableBootstrapMassive2014}, leveraging\citep{2015Leveraging}, mean log-likelihood\citep{Faming2013A} and subsample-based MCMC\citep{2016AB}. Take Bags of Little Bootstrap as an example, \citeauthor{kleinerScalableBootstrapMassive2014}(\citeyear{kleinerScalableBootstrapMassive2014}) combined bootstrap with subsampling, which addressed the high computation problem of bootstrap in big data occasions. As for divide-and-conquer, its most simple type, also called na\"ive-divide-and-conquer or one-shot, is gaining much popularity. Its main idea is dividing complete datasets into several blocks, obtaining the local estimators on each block and calculating the global one by simple averaging. Na\"ive-dc has many applications in regression. For example, to solve large-scale empirical risk minimization problem, \citeauthor{zhangCommunicationefficientAlgorithmsStatistical2012}(\citeyear{zhangCommunicationefficientAlgorithmsStatistical2012}) analyzed a communicational efficient average mixture algorithm based on na\"ive-dc which is robust to the amount of parallelization. \citeauthor{chenSplitandconquerApproachAnalysis2014}(\citeyear{chenSplitandconquerApproachAnalysis2014}) designed a split-and-conquer penalized regression, including $L_1$ penalty, SCAD and MCP. Their estimator is asymptotically equivalent to penalized estimator using entire datasets. Furthermore, some research also applied na\"ive-dc on high-dimensional big datasets. \citeauthor{2015Communication}(\citeyear{2015Communication}) devised an averaging debiased LASSO estimator that can be obtained distributedly and communication efficient in high dimensional settings. \citeauthor{batteyDistributedTestingEstimation2018}(\citeyear{batteyDistributedTestingEstimation2018}) cast their eyes on hypothesis testing. They utilized divide-and-conquer in the context of sparse high dimensional Rao and Wald test, solving computation complexity increase due to high dimensional regularization problems. We can also find some non-linear applications like ridge regression\citep{2013Divide}, SVM\citep{2018Divide} etc. More na\"ive-dc related literature can be seen in \citeauthor{2011Aggregated}(\citeyear{2011Aggregated}), \citeauthor{2016A}(\citeyear{2016A}),\citeauthor{zhuLeastSquareApproximationDistributed2021}(\citeyear{zhuLeastSquareApproximationDistributed2021}). However, complete datasets are necessary for na\"ive-dc before analyzing to find the appropriate block number. While this issue might not arise for conventional big datasets, it is a crucial challenge for big data streams since the latter are incessantly generated with significant volume and velocity, making it infeasible to obtain complete data.

Different from divide-and-conquer, online updating, aiming at big data streams analysis, has no requirement for storing historical data. Specifically, it uses key features extracted from new arriving data batches to continuously update historical data thus is able to achieve computational efficiency and minimal intensive storage. According to \citeauthor{schifanoOnlineUpdatingStatistical2016}(\citeyear{schifanoOnlineUpdatingStatistical2016}), they are the first group to investigate statistical inference on online updating scenario. Their online updating framework is able to solve the predictive residual tests in linear regression and also the estimating equation setting. However, this algorithm needs to start from scratch if new predictive variables are produced with data streams generating. To overcome this disadvantage, \citeauthor{wangOnlineUpdatingMethod2018}(\citeyear{wangOnlineUpdatingMethod2018}) adjusted the online updating framework by bias correction.  Through the use of nested models\citep{1995The,Clogg1995Statistical} in regression coefficients comparison,their method improved estimation efficiency under a variety of practical situations. \citeauthor{leeOnlineUpdatingMethod2020}(\citeyear{leeOnlineUpdatingMethod2020}) extended aforementioned work from the perspective of biased predictive variables in big data streams being corrected afterward under the linear model framework. However, their method is still biased if not knowing the exact data adjusted point. Recently, online updating has also been incorporated in other models for big data analysis, such as online updating GIC for variable selection in generalized linear model\citep{xueOnlineUpdatingInformation2021}, Cox proportional hazards model in online updating framework\citep{wuOnlineUpdatingSurvival2021}, online nonparametric method for functional data\citep{fangyaoOnlineEstimationFunctional2021} etc. To conclude, online updating is accessible to not only ordinary big datasets, but also big data streams, thereby providing a wider range of applications than na\"ive-dc. Nevertheless, most existing online updating frameworks are combined with linear regression, which are known to be sensitive to outliers and heavy-tailed distributions. To address this issue, novel online updating frameworks that are designed to achieve robustness to outliers are urgently needed.

Talking about the outlier problem, one popular solution is quantile regression\citep{1978Regression}. It studies the conditional distribution of each quantile of response variables, thus overcoming the outlier problem and being able to explain potential heteroscedasticity of the datasets. Inspired by its good performance, loads of research incorporate quantile regression into big data analysis framework, especially in divide-and-conquer. Take \citeauthor{chenQuantileRegressionMemory2019}(\citeyear{chenQuantileRegressionMemory2019}) as an example, they used a kernel function to smooth quantile loss function and operated divide-and-conquer iteratively to obtain a linear quantile regression estimator, handling with the constraint of machine number in na\"ive-dc. Moreover, \citeauthor{huDistributedQuantileRegression2021}(\citeyear{huDistributedQuantileRegression2021}) addressed the nondifferentiable problem of quantile regression by developing CSL\citep{jordanCommunicationEfficientDistributedStatistical2019} method to speed up big data algorithm. We can also find other research combining quantile regression with subsampling and online updating like \citeauthor{chenQuantileRegressionBig2020}(\citeyear{chenQuantileRegressionBig2020}) and \citeauthor{wangRenewableQuantileRegression2022}(\citeyear{wangRenewableQuantileRegression2022}). By using online updating, Wang's(\citeyear{wangRenewableQuantileRegression2022}) method can be applied to big data streams. Particularly, \citeauthor{wangRenewableQuantileRegression2022}(\citeyear{wangRenewableQuantileRegression2022}) analyzed quantile regression from the view of maximum likelihood and incorporated it in online updating algorithm. Without operations in check loss function, it is more convenient than the work of \citeauthor{chenQuantileRegressionMemory2019}(\citeyear{chenQuantileRegressionMemory2019}), and \citeauthor{chenQuantileRegressionBig2020}(\citeyear{chenQuantileRegressionBig2020}). Some variants of quantile regression like composite quantile regression and adaptive quantile regression are also investigated in big data analysis, which can be seen in \citeauthor{jiangCompositeQuantileRegression2018}(\citeyear{jiangCompositeQuantileRegression2018}),\citeauthor{jiangAdaptiveQuantileRegressions2021}(\citeyear{jiangAdaptiveQuantileRegressions2021}) for details.

\subsection{Motivations}

While quantile regression is investigated thoroughly, other members in M-estimation family, such as Huber regression\citep{Huber1964Robust}, have received litte attention. Huber regression transforms the loss of big residuals from quadratic form into linear form to cut down weights, thus achieving good robustness to outliers. Therefore, Huber regression presents a feasible alternative to quantile regression in solving outlier-related problems. Regrettably, there has been little research on the use of Huber regression in big data analysis. To date, the only literature used Huber loss in big data regression is Luo's(\citeyear{luoDistributedAdaptiveHuber2022}) work. They embedded Huber loss function in Jordan's CSL\citep{jordanCommunicationEfficientDistributedStatistical2019} framework, adding two data-driven parameters in Huber regression to achieve a balance of statistical optimization and communication efficiency. Nonetheless, there is no literature focusing on combining Huber regression with online updating to solve analysis in big data streams.

In order to solve the storage limitation and outlier problems aforementioned in big data streams analysis, we propose a novel Online Updating Huber Regression algorithm, which is inspired by the maximum likelihood idea in Wang's(\citeyear{wangRenewableQuantileRegression2022}) online updating algorithm. The resulting online updating estimator in our algorithm is renewed by current data and summary statistics of historical data, which is computationally efficient. Specifically, by assuming local estimators generated from a multivariate normal distribution and analyzing it from the view of maximum likelihood, a weighted least square type objective function can be obtained. In the objective function, two key features of each batch can be extracted, which can be updated and easily operated in big data streams. Meanwhile, we prove that the proposed estimator is asymptotically equivalent to Oracle estimator using the entire dataset and has lower computation complexity compared with Oracle one. Numerical experiments on both simulation and real data are also included to verify the theoretical results and illustrate the good performance of our new method. 

Our study makes two primary contributions. Firstly, we employ online updating instead of the widely used divide-and-conquer method to address the limited computer memory constraint so that it can be applied on the analysis of big data streams. Furthermore, by integrating Huber regression into online updating framework, we broaden the usage of online updating in robust regression other than pure linear regression. The second contribution is about Huber regression. Based on the state-of-art in big data analysis using loss functions in M-estimation, we adopt Huber regression as an alternative to quantile regression to increase robustness. It not only fills certain research gaps in Huber regression, but also studies feasibility of generalizing quantile regression to other loss functions in M-estimation.

The rest of this paper is organized as follows. Section 2 explains the methodology of our Online Updating Huber Regression by obtaining the main online updating estimator. Section 3 proves some good properties of the Online Updating Huber Regression estimator, including asymptotic equivalency and low computation complexity. Section 4 introduces simulation and real data analysis results in order to verify the good performance of our algorithm. Section 5 is about conclusions and future work of our research.

\section{Methodology}

\subsection{Notation and Oracle Estimator}
Suppose there are $N_b$ identically independent distributed samples $\{(y_n,\textbf{x}_n),n=1,2,...,N_b\}$. $y\in \rm{R}$ is response variable and $\textbf{x} = (x^1,x^2,...,x^p)^T \in \textbf{R}^{p\times 1}$ is p-dimensional covariates. In the big data streams scenario, suppose $N_b$ samples are generated in form of $b$ data subsets $\{D_1,D_2,...,D_b\}$ sequentially. $D_t=\{(y_{ti},\textbf{x}_{ti}),i=1,2,...,n_t\}$ stands for the $t$-th subset. Our aim is to fit a model $y_{ti}={\textbf{x}_{ti}}^T \bm{\theta} + {\epsilon_{ti}}$, where  $\bm{\theta}=(\theta_1,\theta_2,...,\theta_p)^T$ is a p-dimensional unknown parameter with true value $\bm{\theta}_0=(\theta_{01},\theta_{02},...,\theta_{0p})^T$, and $\epsilon_{ti}$ is random error which can be heavy-tailed distributed, heteroscedastic distributed or contaminated by outliers. $f(\cdot)$ is the probability density distribution of $\epsilon_{ti}$.

As we want to prove that our Online Updating estimator is asymptotic equivalent to Oracle one using entire dataset follow-up, we first point the expression of Oracle Huber Regression estimator here. According to Huber(\citeyear{Huber1964Robust}), the expression of Oracle Huber Regression estimator of $\bm{\theta}_0$ is as follows:

\begin{equation}\label{oracle}
	{\widehat{\bm{\theta}}_{N_b}=\text{arg}\min_{\bm{\theta}}Q_{N_b}(\bm{\theta}),\ \ \ Q_{N_b}(\bm{\theta})=\frac{1}{N_b}\sum_{t=1}^b\sum_{i=1}^{n_t}\rho(y_{ti}-\textbf{x}_{ti}^T\bm{\theta})}
\end{equation}
where $\rho(u)=\frac{u^2}{2}I(|u|<k)+\{k|u|-\frac{k^2}{2}\}I(|u|\geq k)$.

Moreover, under following regularity conditions \cite{Huber1964Robust,Huber1981,wangRenewableQuantileRegression2022}:

(N1) For $\lambda(\bm{\theta})=E(\textbf{x}\psi(y-\textbf{x}^T\bm{\theta}))$, where $\psi(\cdot)=\rho^{'}(\cdot)$, $\bm{\theta}_0$, the true value of $\bm{\theta}$, satisfies $\lambda(\bm{\theta}_0)=0$.

(N2)$E(\psi^2)<\infty,0<E(\psi^{'})<\infty$.

(N3) Covariates $\textbf{x}_{ti}$ satisfies $\frac{max_{t=1,...,b,i=1,...,n_t \Vert {\textbf{x}_{ti}} \Vert}}{\sqrt{N_b}}\rightarrow\infty,\Vert {\textbf{x}_{ti}} \Vert=({\textbf{x}_{ti}^T\textbf{x}_{ti}})^{\frac{1}{2}}$.

(N4)Exist a positive matrix $\bm{\Sigma}$ satisfying $\frac{1}{N_b}\sum_{t=1}^b\sum_{i=1}^{n_t}\textbf{x}_{ti}\textbf{x}_{ti}^T\xrightarrow{p}\bm{\Sigma}, N_b\rightarrow\infty$, where $\xrightarrow{p}$ stands for convergence in probability.

We can obtain the asymptotic normal distribution of Oracle Huber Regression estimator $\widehat{\bm{\theta}}_{N_b}$ as follows:
	
\begin{equation}\label{oracle-normal}
	{\sqrt{N_b}\Big(\widehat{\bm{\theta}}_{N_b}-\bm{\theta}_0\Big)\overset{d}\rightarrow N\Big(0,\frac{E(\psi^2)}{[E(\psi')]^2}\bm{\Sigma^{-1}} \Big)},\\\
	N_b\rightarrow\infty
\end{equation}

where $\frac{E(\psi^2)}{[E(\psi')]^2}$ is a function of $f(\cdot)$, $\xrightarrow{d}$ stands for convergence in distribution.

\subsection{Online Updating Huber Robust Regression Estimator}

In this part, we mainly deduce the expression of Online Updating Huber Robust Regression (UHR) estimator. Firstly, we can calculate the local Huber estimator in each subset. Suppose $\widehat{\bm\theta}_t$ is the local Huber estimator in the $t$-th subset. According to Huber(\citeyear{Huber1964Robust}), its formula is like Eq.(\ref{local}).

\begin{equation}\label{local}
	{\widehat{\bm{\theta}}_{t}=\text{arg}\min_{\bm{\theta}}Q_{t}(\bm{\theta}),\ \ \ Q_{t}(\bm{\theta})=\frac{1}{n_t}\sum_{i=1}^{n_t}\rho(y_{ti}-\textbf{x}_{ti}^T\bm{\theta})}
\end{equation}

As the entire dataset is divided into several subsets, subset size $n_t$ can be much smaller than total sample size $N_b$. Therefore, the calculation of Eq.(\ref{local}) can be more computationally efficient and less storage intensive than Eq.(\ref{oracle}), so that we can solve it faster. Furthermore, under regularity conditions (N1)-(N4), local estimator $\widehat{\bm{\theta}}_{t}$ follows asymptotic normal distribution in Eq.(\ref{local_dis}), where $\bm{\Sigma}_t=E(\textbf{x}_{ti}\textbf{x}_{ti}^T)$.
\begin{equation}\label{local_dis}
	{\sqrt{n_t}\Big(\widehat{\bm{\theta}}_{t}-\bm{\theta}_0\Big)\overset{d}\rightarrow N\Big(\textbf{0},\frac{E(\psi^2)}{[E(\psi')]^2}\bm{\Sigma^{-1}}_t \Big)},\\\
	n_t\rightarrow\infty
\end{equation}

Taking the idea in Wang's(\citeyear{wangRenewableQuantileRegression2022}) work as an inspiration, we consider the local estimators from the perspective of maximum likelihood. Due to identical independent distribution among total samples, local estimator $\widehat{\bm{\theta}}_1,...,\widehat{\bm{\theta}}_b$ are mutually independent. Thus, we can take $\{\widehat{\bm{\theta}}_1,...,\widehat{\bm{\theta}}_b\}$ as a sample generated from the following multivariate normal distribution.
\begin{equation}\label{multi-normal}
	\Big\{N\Big(\bm{\theta}_0,\frac{1}{n_1}\frac{E(\psi^2)}{[E(\psi')]^2}\bm{\Sigma}_1^{-1}\Big),\cdots, N\Big(\bm{\theta}_0,\frac{1}{n_b}\frac{E(\psi^2)}{[E(\psi')]^2}\bm{\Sigma}_b^{-1}\Big)\Big\}
\end{equation}
Based on Eq.(\ref{multi-normal}), the maximum likelihood function of $\bm{\theta}$ is as follows.
\begin{equation}\label{max-likeli}
	L(\bm{\theta})=\prod_{t=1}^b\big{(}\frac{1}{\sqrt{2\pi}}\big{)}^p\big{(}\frac{1}{n_t}\frac{E(\psi^2)}{[E(\psi')]^2}|\bm{\Sigma}_t^{-1}|\big{)}^{-\frac{1}{2}}{\rm exp}\big{\{}-\frac{n_t}{2}\frac{[E(\psi')]^2}{E(\psi^2)} (\widehat{\bm{\theta}}_t-\bm{\theta})^T\bm{\Sigma}_t(\widehat{\bm{\theta}}_t-\bm{\theta})\big{\}}
\end{equation}
By turning Eq.(\ref{max-likeli}) into logarithmic form, we can get Eq.(\ref{log-like})

\begin{equation}\label{log-like}
	{\rm{log}}(L(\bm{\theta}))=C-\frac{1}{2}\frac{[E(\psi')]^2}{E(\psi^2)}\sum_{t=1}^{b}n_t(\widehat{\bm{\theta}}_t-\bm{\theta})^T\bm{\Sigma}_t(\widehat{\bm{\theta}}_t-\bm{\theta})
\end{equation}
where $C$ is a constant, $\frac{[E(\psi')]^2}{E(\psi^2)}$ is a function of $f(\cdot)$, which is irrelevant with $\bm{\theta}$ as well. 
By maximizing Eq.(\ref{log-like}), we can obtain the UHR estimator. Meanwhile, base on non-negativity of $\frac{[E(\psi')]^2}{E(\psi^2)}$, maximizing Eq.(\ref{log-like}) is equivalent to minimize the following loss function.

\begin{equation}\label{loss-func-1}
	\sum_{t=1}^{b}n_t(\widehat{\bm{\theta}}_t-\bm{\theta})^T\bm{\Sigma}_t(\widehat{\bm{\theta}}_t-\bm{\theta})
\end{equation}

By using $\frac{1}{n_t}\sum_{i=1}^{n_t}\textbf{x}_{ti}\textbf{x}_{ti}^T$ to replace $\bm{\Sigma}_t,t=1,...,b$ according to $\bm{\Sigma}_t=E(\textbf{x}_{ti}\textbf{x}_{ti}^T)$, we can obtain the final objective function as follows, where $\textbf{X}_t=(\textbf{x}_{t1},...,\textbf{x}_{tn_t})^T$.
\begin{equation}\label{loss-func-final}
	Q(\bm{\theta})=\sum_{t=1}^{b}(\widehat{\bm{\theta}}_t-\bm{\theta})^T\textbf{X}_t^T\textbf{X}_t(\widehat{\bm{\theta}}_t-\bm{\theta})
\end{equation}
 Thus Online Updating Huber Robust Regression estimator of $\bm{\theta}_0$, $\widehat{\bm{\theta}}_{N_b}^{uhr}$, can be obtained by minimizing Eq.(\ref{loss-func-final}).

\begin{equation}\label{updating_estimator}
    \begin{split}
    \widehat{\bm{\theta}}_{N_b}^{uhr}&=\text{arg}\min_{\bm\theta}Q(\bm\theta)\\
	&=\Big(\sum_{t=1}^b\textbf{X}_t^T\textbf{X}_t\Big)^{-1}\Big(\sum_{t=1}^b\textbf{X}_t^T\textbf{X}_t\widehat{\bm\theta_t}\Big)
    \end{split}
\end{equation}

According to  Eq.(\ref{updating_estimator}), we can only extract two key features from each subset, i.e. $\textbf{X}_t^T\textbf{X}_t$ and $\widehat{\bm\theta}_t$, with no need to store entire dataset. This enables us to save computer memory and speed up computation process, thus more available to big data streams.

\begin{figure}
	\centerline{\includegraphics[height=6.5cm]{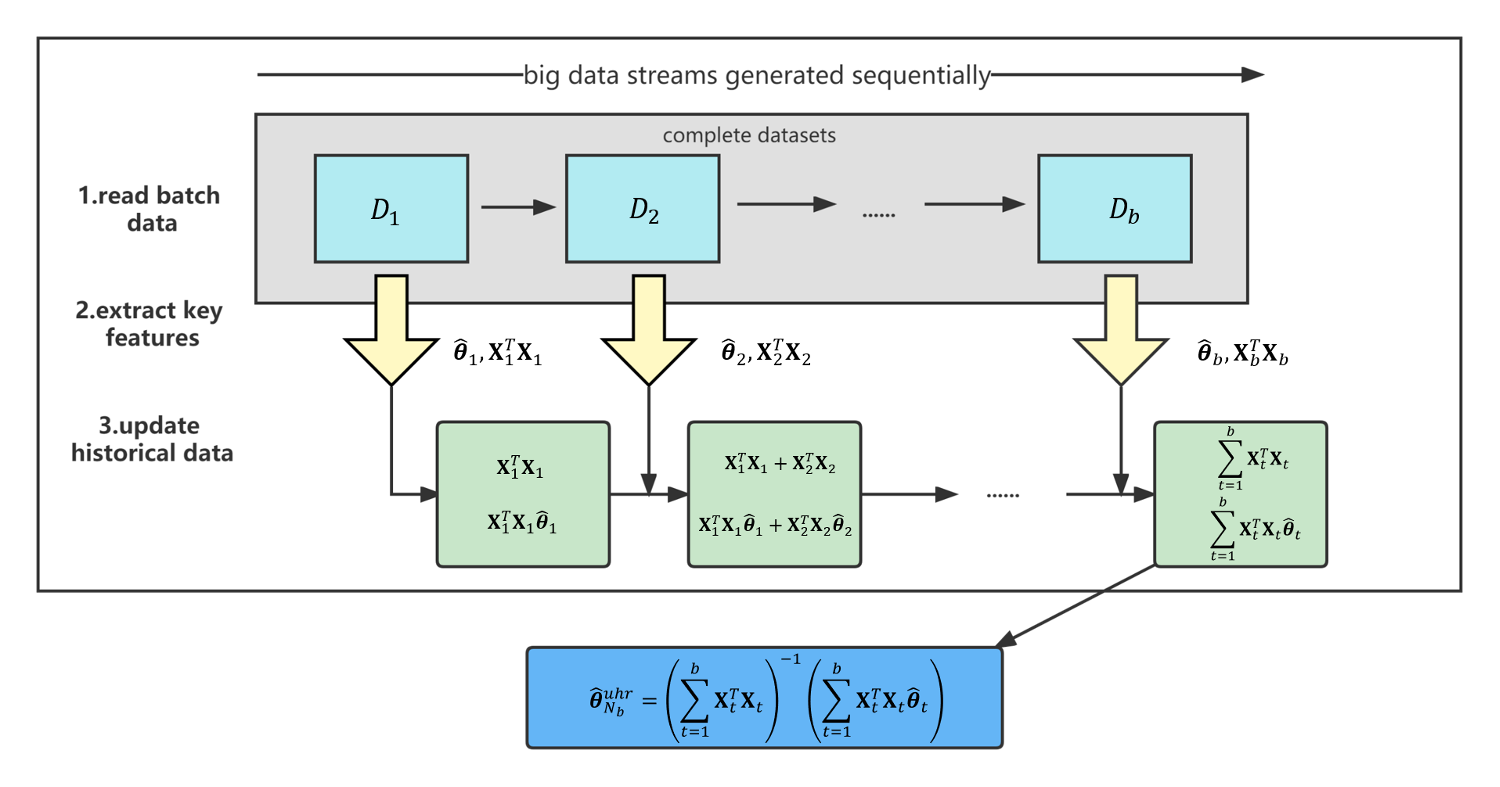}}
	\caption{Online Updating algorithm for Huber regression estimator}
	\label{algorithm}
\end{figure}

The UHR algorithm implementation process can be seen in Figure \ref{algorithm}. To start the algorithm, we import the first subset $D_1=\{(y_{1i},\textbf{x}_{1i}),i=1,...,n_1\}$. By using function \textbf{rlm} of package \textbf{MASS} in R, the local estimator of the first subset $\widehat{\bm\theta}_1$ can be calculated. Then two features in the first subset $\textbf{X}_1^T\textbf{X}_1$ and $\textbf{X}_1^T\textbf{X}_1\widehat{\bm\theta}_1$ can be obtained. For the second subset $D_2=\{(y_{2i},\textbf{x}_{2i}),i=1,...,n_2\}$, use the same method to calculate $\widehat{\bm\theta}_2$ and $\textbf{X}_2^T\textbf{X}_2$, then add them into $\textbf{X}_1^T\textbf{X}_1$ and $\textbf{X}_1^T\textbf{X}_1\widehat{\bm\theta}_1$. The sum obtained, $\textbf{X}_1^T\textbf{X}_1+\textbf{X}_2^T\textbf{X}_2$ and $\textbf{X}_1^T\textbf{X}_1\widehat{\bm\theta}_1+\textbf{X}_2^T\textbf{X}_2\widehat{\bm\theta}_2$ are new results of subset $D_1$ updated by subset $D_2$. The rest subsets can be processed in the same manner. After the last subset $D_b=\{(y_{bi},\textbf{x}_{bi}),i=1,...,n_1\}$ is imported, we can obtain key elements, i.e. $\sum_{t=1}^b\textbf{X}_t^T\textbf{X}_t$ and $\sum_{t=1}^b\textbf{X}_t^T\textbf{X}_t\widehat{\bm\theta_t}$, and finally obtain the Online Updating estimator $\widehat{\bm{\theta}}_{N_b}^{uhr}$ updated by the entire dataset. The pseudo-code can be seen in Algorithm 1.

\begin{algorithm}[t]
	\caption{Online updating algorithm for Huber regression estimator} 
	\hspace*{0.02in} {\bf Input:} 
	input subsets $D_t=\{(y_{ti},\textbf{x}_{ti}), i=1,2,...,n_t\}$ \\
	\hspace*{0.02in} {\bf Output:} 
	Online Updating Huber Estimator $\widehat{\bm\theta}_{N_b}^{uhr}$
	\begin{algorithmic}[1]
		\State initialize $\textbf{U}_t=\bm{0}_{p\times p}, \textbf{V}_t = 0$
		\For{t=1,2,...,b} 
		\State $\widehat{\bm{\theta}}_{t}=\text{arg}\min_{\bm{\theta}}Q_{t}(\bm{\theta}),\ \ \ Q_{t}(\bm{\theta})=\frac{1}{n_t}\sum_{i=1}^{n_t}\rho(y_{ti}-\textbf{x}_{ti}^T\bm{\theta})$
		\State $\textbf{U}_t = \textbf{U}_t + \textbf{X}_t^T\textbf{X}_t, \textbf{V}_t=\textbf{V}_t+\textbf{X}_t^T\textbf{X}_t\widehat{\bm\theta_t}$
		\EndFor
		\State \Return $\widehat{\bm\theta}_{N_b}^{uhr}=\textbf{U}_t^{-1}\textbf{V}_t$
	\end{algorithmic}
\end{algorithm}

\section{Properties of the Online Updating estimator}

\subsection{Asymptotic Normality}
In order to prove the asymptotic normality of our UHR estimator, we first add some assumptions into regularity conditions (N1)-(N4)\citep{wangRenewableQuantileRegression2022}.

(N5) Take $n=\frac{N_b}{b}$ as the average size of each subset. Suppose $\frac{n}{\sqrt{N_b}}\rightarrow\infty$, and all $n_t$ diverge in the same order $O(n)$, i.e., $c_1\leq {\rm{min}}_t n_t/n\leq {\rm{max}}_t n_t/n \leq c_2$ for some positive constants $c_1$ and $c_2$.

(N6) Exist positive definite matrix $\bm{\Sigma}_1,...,\bm{\Sigma}_b$ which satisfy $\frac{1}{n_t}\sum_{i=1}^{n_t}\textbf{x}_{ti}\textbf{x}_{ti}^T\xrightarrow{p}\bm{\Sigma}_t, n_t\rightarrow\infty$ and converge to $\bm{\Sigma}_t$ in the same rate, i.e. convergence rate is irrelevant with $t$.

Obviously, $\frac{1}{n_t}\sum_{i=1}^{n_t}\textbf{x}_{ti}\textbf{x}_{ti}^T-\bm{\Sigma}_t=\textbf{O}(\frac{1}{\sqrt{n_t}})$ due to $\bm{\Sigma}_t=E(\textbf{x}_{ti}\textbf{x}_{ti}^T)$. Moreover, according to (N5) and (N6), all subsets have the same convergence rate, which is $\textbf{O}(\frac{1}{\sqrt{n_t}})=\textbf{O}(\frac{1}{\sqrt{n}})$.

Under regularity conditions (N1)-(N6), UHR estimator $\widehat{\bm{\theta}}_{N_b}^{uhr}$ follows asymptotic distribution as follows:

\begin{equation}\label{asymptotic-normal}
	[\Phi(f)]^{-\frac{1}{2}}\sqrt{N_b}(\widehat{\bm\theta}_{N_b}^{uhr}-\bm{\theta}_0)\overset{d}\rightarrow N(\bm{0},\bm{I}_p),
\end{equation}
where $\Phi(f)=\frac{E(\psi^2)}{[E(\psi')]^2}\Big(\sum_{t=1}^b \frac{n_t}{N_b}\bm{\Sigma}_t\Big)^{-1}$. The proof is as follows. 

For local Huber estimator $\bm{\widehat{\theta}}_t$, it follows asymptotic condition:

\begin{equation}\label{proof-1}
	\widehat{\bm{\theta}}_t-\bm{\theta}_0 = \bm{\Sigma}_t^{-1}{\frac{1}{E(\psi^{'})}}{\frac{1}{n_t}}\sum_{i=1}^{n_t}\textbf{x}_{ti}\psi(\epsilon_{ti})+O_p\Big(\frac{1}{n_t}\Big)
\end{equation}
where $\psi(\epsilon_{ti})=\rho^{'}(\epsilon_{ti})=\left\{
\begin{array}{ll}
	-k, &\epsilon_{ti}<-k \\
	\epsilon_{ti}, &\lvert\epsilon_{ti}\lvert\leq k \\
	k, &\epsilon_{ti}>k
\end{array}.
\right.$

According to Eq.(\ref{updating_estimator}), the relationship with $\widehat{\bm{\theta}}_{N_b}^{uhr}$ and $\widehat{\bm{\theta}}_t$ is

\begin{equation}\label{proof-2}
	\widehat{\bm{\theta}}_{N_b}^{uhr}=\Big(\sum_{t=1}^{b} w_{t}\frac{\textbf{X}_t^T\textbf{X}_t}{n_t}\Big)^{-1}\Big(\sum_{t=1}^{b} w_{t}\frac{\textbf{X}_t^T\textbf{X}_t\widehat{\bm\theta_t}}{n_t}\Big)
\end{equation}
where $w_{t}=\frac{n_t}{N_b}$.
By direct calculation, we can obtain that

\begin{equation}\label{proof-3}
	\sqrt{N_b}(\widehat{\bm{\theta}}_{N_b}^{uhr}-\bm{\theta}_0)=\Big(\sum_{t=1}^{b} w_{t}\frac{\textbf{X}_t^T\textbf{X}_t}{n_t}\Big)^{-1}\Big(\sqrt{N_b}\sum_{t=1}^{b} w_{t}\frac{\textbf{X}_t^T\textbf{X}_t}{n_t}(\widehat{\bm\theta_t}-\bm{\theta}_0)\Big)
\end{equation}
By $\sqrt{N_b}\Bigg(\sum_{t=1}^{b} w_{t}\Big(\frac{\textbf{X}_t^T\textbf{X}_t}{n_t}-\bm{\Sigma}_t\Big)(\widehat{\bm\theta_t}-\bm{\theta}_0)\Bigg)=O_p\Big(\frac{b}{\sqrt{N_b}}\Big)$
and regularity conditions (N1)-(N6), we can obtain

\begin{equation}\label{proof-4}
	\begin{aligned}
		&\sqrt{N_b}\Big(\sum_{t=1}^{b} w_{t}\frac{\textbf{X}_t^T\textbf{X}_t}{n_t}(\widehat{\bm\theta_t}-\bm{\theta}_0)\Big)\\
		=&\sqrt{N_b}\Big(\sum_{t=1}^{b}w_t\bm{\Sigma}_{t}(\widehat{\bm{\theta}}_t-\bm{\theta}_0)\Big)+ \sqrt{N_b}\Bigg(\sum_{t=1}^{b} w_{t}\Big(\frac{\textbf{X}_t^T\textbf{X}_t}{n_t}-\bm{\Sigma}_t\Big)(\widehat{\bm\theta_t}-\bm{\theta}_0)\Bigg)\\
		=&\sqrt{N_b}\Big(\sum_{t=1}^{b}w_t\bm{\Sigma}_{t}\Big(\bm{\Sigma}_t^{-1}{\frac{1}{E(\psi^{'})}}{\frac{1}{n_t}}\sum_{i=1}^{n_t}\textbf{x}_{ti}\psi(\epsilon_{ti})+O_p\Big(\frac{1}{n_t}\Big)\Big) + O_p\Big(\frac{b}{\sqrt{N_b}}\Big)\\
		=&\frac{1}{\sqrt{N_b}}\sum_{t=1}^{b}{\frac{1}{E(\psi^{'})}}\sum_{i=1}^{n_t}\textbf{x}_{ti}\psi(\epsilon_{ti})+O_p\Big(\frac{b}{\sqrt{N_b}}\Big)
	\end{aligned}
\end{equation}
As $\sum_{t=1}^{b}w_t=1$, we have
\begin{equation}\label{proof-5}
	\sum_{t=1}^{b}w_t\frac{\textbf{X}_t^T\textbf{X}_t}{n_t}-\sum_{t=1}^{b}w_t\bm{\Sigma}_t=\sum_{t=1}^{b}w_t\Big(\frac{\textbf{X}_t^T\textbf{X}_t}{n_t}-\bm{\Sigma}_t\Big)=o_p(1)
\end{equation}
Then we can obtain that

\begin{equation}\label{proof-6}
	\begin{aligned}
		&[\Phi(f)]^{-\frac{1}{2}}\sqrt{N_b}(\widehat{\bm\theta}_{N_b}^{uhr}-\bm{\theta}_0)\\
		= &[\Phi(f)]^{-\frac{1}{2}}\Big(\sum_{t=1}^{b} w_{t}\frac{\textbf{X}_t^T\textbf{X}_t}{n_t}\Big)^{-1}\Big(\sqrt{N_b}\sum_{t=1}^{b} w_{t}\frac{\textbf{X}_t^T\textbf{X}_t}{n_t}(\widehat{\bm\theta_t}-\bm{\theta}_0)\Big)\\
		= &[\Phi(f)]^{-\frac{1}{2}}\Big(\sum_{t=1}^{b} w_{t}\frac{\textbf{X}_t^T\textbf{X}_t}{n_t}\Big)^{-1}\Big(\frac{1}{\sqrt{N_b}}\sum_{t=1}^{b}{\frac{1}{E(\psi^{'})}}\sum_{i=1}^{n_t}\textbf{x}_{ti}\psi(\epsilon_{ti})+O_p\Big(\frac{b}{\sqrt{N_b}}\Big)\Big)\\
		&\xrightarrow{d}N(\textbf{0},\bm{I}_p)
	\end{aligned}
\end{equation}

If the covariates of each batch are homogeneous, i.e., $\Sigma_1=\cdots=\Sigma_b=\Sigma$, the $\Phi(f)=\frac{E(\psi^2)}{[E(\psi')]^2}\bm{\Sigma}^{-1}$, which coincides the asymptotic variance of oracle estimator in Eq.(\ref{oracle-normal}). The updating estimator $\widehat{\bm\theta}_{N_b}^{uhr}$ is asymptotically equivalent with the oracle estimator $\widehat{\bm\theta}_{N_b}$.

\subsection{Computation Complexity}
In this part, we do a rough calculation on the computation complexity of Oracle Huber Regression and Online Updating Huber Regression, attesting our Online Updating algorithm has a lower computation complexity.

In our work, we use function \textbf{rlm} in \textbf{MASS} package of R to calculate the local estimator and Oracle estimator obtained by entire dataset. Function \textbf{rlm} is implemented by Iteratively Reweighted Least Squares. For the sake of calculation, suppose \textbf{rlm}'s maximum iterative time is $K$, total sample size is $N$ and is divided into $b$ subsets evenly with $n$ samples each.
Therefore, the computation complexity of solving the local estimator and the Oracle Huber estimator are about $O(K(pn^2+pn+np^3))$ and $O(KpN^2+(Kp+Kp^2)N+Kp^3)$ respectively. As Online Updating algorithm contains $b$ loops and calculating the key features of each subset cost about $O(p^2n)$, total computation complexity of Online Updating algorithm is about $O(Kb(pn^2+pn+n+p^3)+bp^2n)$. By $N=nb$, we can simplify it into $O(KpNn+(Kp^2+Kp+1)N+Kbp^3)$. Although $K$ is set with a large number(mostly 1000), it can achieve convergence within 10 times of iteration in practice, while total sample $N$ can reach a million or more, thus $K\ll N$. As for $p$, we only consider low-dimension cases where $p\ll N$. To sum up, the dominants part of computation complexity of Oracle and Online Updating ones are $O(KpN^2)$ and $O(KpNn)$. By thorough comparison, we can find the only difference between these two are $N$ and $n$, where $n \ll N$. In conclusion, theoretically, our Online Updating Huber Regression has lower computation complexity than Oracle Huber Regression.

\section{Numerical Studies}

In this section, we demonstrate the performance of our UHR algorithm by comparing it with other 4 algorithms in 6 different synthetic datasets. We also apply UHR on a real airline dataset released by American Statistical Association, proving it is able to solve real problems as well. 

\subsection{Simulations}

\subsubsection{Experiment Settings}
In simulations, we generate data using the following model:
$$y_{ti}=\textbf{x}_{ti}^T\bm{\theta}+h(\textbf{x}_{ti})\epsilon_{ti}$$
where coefficients $\bm{\theta}=(\theta_1,\theta_2,\theta_3,\theta_4)^T=(1,-1,2,-2)^T$, covariates $\textbf{x}_{ti}$ follow a multivariate normal distribution with zero mean and independent type of covariance matrix. Function $h(\textbf{x}_{ti})$ is used for controlling the homoscedasticity or heteroscedasticity of random error. Specifically, if random error is homogeneous, $h(\textbf{x}_{ti})=1$, else $h(\textbf{x}_{ti})=\sum_{j=1}^{4}\textbf{x}_{ti}^{j}$. For random error $\epsilon_{ti}$, consider the following 5 cases:

(1) $\epsilon_{ti}\sim N(0,1)$;

(2) $\epsilon_{ti}\sim t(3)$;

(3) $\varepsilon_{ti}\sim t(3)$ with $15\%$ of response $y_{ti}$ replaced by $y_{ti}+20$;

(4) $\epsilon_{ti}\sim 0.85 N(0,1)+0.15N(0,8)$; 

(5) $\epsilon_{ti}\sim {\rm{Cauchy}} (0,1)$.

Besides our Online Updating Huber Regression algorithm(UHR), we choose other 4 comparing algorithms, which are Ordinary Least Regression(OLS), Renewable Least Regression\citep{luoRenewableEstimationIncremental2020a}, Oracle Huber Regression\citep{Huber1964Robust}(OHR), Divide-and-Conquer Huber Regression(DC-HR). OLS and RLS are to prove good performance of our UHR algorithm in robustness. OHR is to investigate whether Online Updating estimator has the same estimate efficiency with Oracle one. DC-HR is a simple combination of na\"ive-dc and Huber regression to compare the differences between divide-and-conquer and online updating methods. Moreover, to reduce randomness, we repeat the whole simulation process 500 times. 

As for evaluation indexes, we judge the results from estimation and calculation efficiency. There are three indexes: $MSE$, $MAE$ and $time$, whose expressions are as Eq.(\ref{mse})$\sim$Eq.(\ref{time}).

\begin{equation}\label{mse}
	MSE(\bm \theta)=\frac{1}{500}\sum_{s=1}^{500}\sum_{j=1}^{4}(\widehat{\bm \theta}_j^{(s)}-\bm{\theta}_j)^2
\end{equation} 

\begin{equation}\label{mae}
	MAE(\bm \theta)=\frac{1}{500}\sum_{s=1}^{500}\sum_{j=1}^{4}|\widehat{\bm \theta}_j^{(s)}-\bm{\theta}_j|
\end{equation} 

\begin{equation}\label{time}
	time=\frac{\sum_{s=1}^{500} elapsed^{(s)}}{500}
\end{equation} 

For illustration, $\hat{\theta_j}^{(s)}$ is the estimator of the $j$-th dimension of $\theta$in the $s$-th repetition. Thus, $MSE$ and $MAE$ are the average of the sum of the mean squared error and mean absolute error of each dimension in 500 repetitions. $time$ is the average cost time of certain random error running 500 times, which use function \textbf{system.time()} in R to record elapsed time.

Our simulations include 4 experiments in total:

\textbf{Experiment\ 1}: Fix total sample size $N_b = 1,000,000$, change subsets number $b = \{100,200,500,1000\}$. This experiment is to compare the differences in estimation and calculation efficiency between different algorithms and also the influences of subsets number $b$ on them.

\textbf{Experiment\ 2}: Fix subset size $n_t= 100$, change subsets number $b=\{10,100,1000,10000\}$. Accordingly, total sample size $N_b$ varies from $1,000$ to $1,000,000$. This experiment is to investigate the changes in estimation efficiency of each algorithm when subset number $b$ increases.

\textbf{Experiment\ 3}: Fix subset size $n_t= 5000$, change total sample size $N_b=\{2.5\times10^5, 5\times10^5, 7.5\times10^5,10\times10^5\}$. Accordingly, subset number $b$ varies from $50$ to $200$. This experiment is to investigate the influence of $N_b$ on each algorithm. We also investigate the relationship between $N_b$ and running time in this experiment to verify computation complexity obtained before from the simulation perspective.

\textbf{Experiment\ 4}: Fix subset size $n_t= 1000$, change total sample size $N_b=\{1\times10^5, 2\times10^5,..., 8\times10^5\}$. Accordingly, subset number $b$ varies from $100$ to $800$. This experiment is very similar to \textbf{Experiment\ 3} but with smaller $n_t$ to investigate the difference between different scales of $n_t$.

\subsubsection{Estimation Efficiency}
In estimation efficiency, we focus on \textbf{Experiment\ 1} and \textbf{Experiment\ 2}. For \textbf{Experiment\ 1}, Table \ref{e1-case1} to Table \ref{e1-case6} present the $MSE$ and $MAE$ under 6 random error distribution. According to Table \ref{e1-case1} to Table \ref{e1-case6}, we can draw the following conclusions:

(1) For heavy-tailed, heterogeneous error and random error with outliers cases, the estimation error of Huber family(OHR, DC-HR, UHR) is much smaller than that of Linear Regression family(OLS, RLS), demonstrating Huber algorithms are more robust than Linear Regression algorithms.

(2) When random error is distributed in case 1 $\sim$ case 4, our UHR has very close estimation errors with OHR. This verifies that Online Updating estimator $\widehat{\bm{\theta}}_{N_b}^{uhr}$ are asymptotic equivalent with Oracle Huber estimator from experiment perspective.

(3) When random error is distributed in case 6, i.e. Cauchy distribution, we can find Linear Regression algorithms are invalid with such big $MSE$ and $MAE$, while Huber algorithms are still performing well. It is worth mentioning that distributed algorithms DC-HR and UHR even have obvious estimation improvements compared with Oracle Huber estimator. This phenomenon may be related to characteristics of Cauchy distribution that variance does not exist. Due to the high volatility of data, using the whole dataset may not come with the best estimating efficiency, while subsets with smaller sizes may perform better.

(4) By comparing two distributed Huber algorithms, we find that in most cases, UHR has smaller estimation errors than DC-HR, which attests that Online Updating algorithm has a better performance than divide-and-conquer. Even though DC-HR can perform better than UHR in some cases, we still prefer UHR due to the wider range of applications for online updating in big data streams.

\begin{figure}
	\centerline{\includegraphics[height=6.5cm]{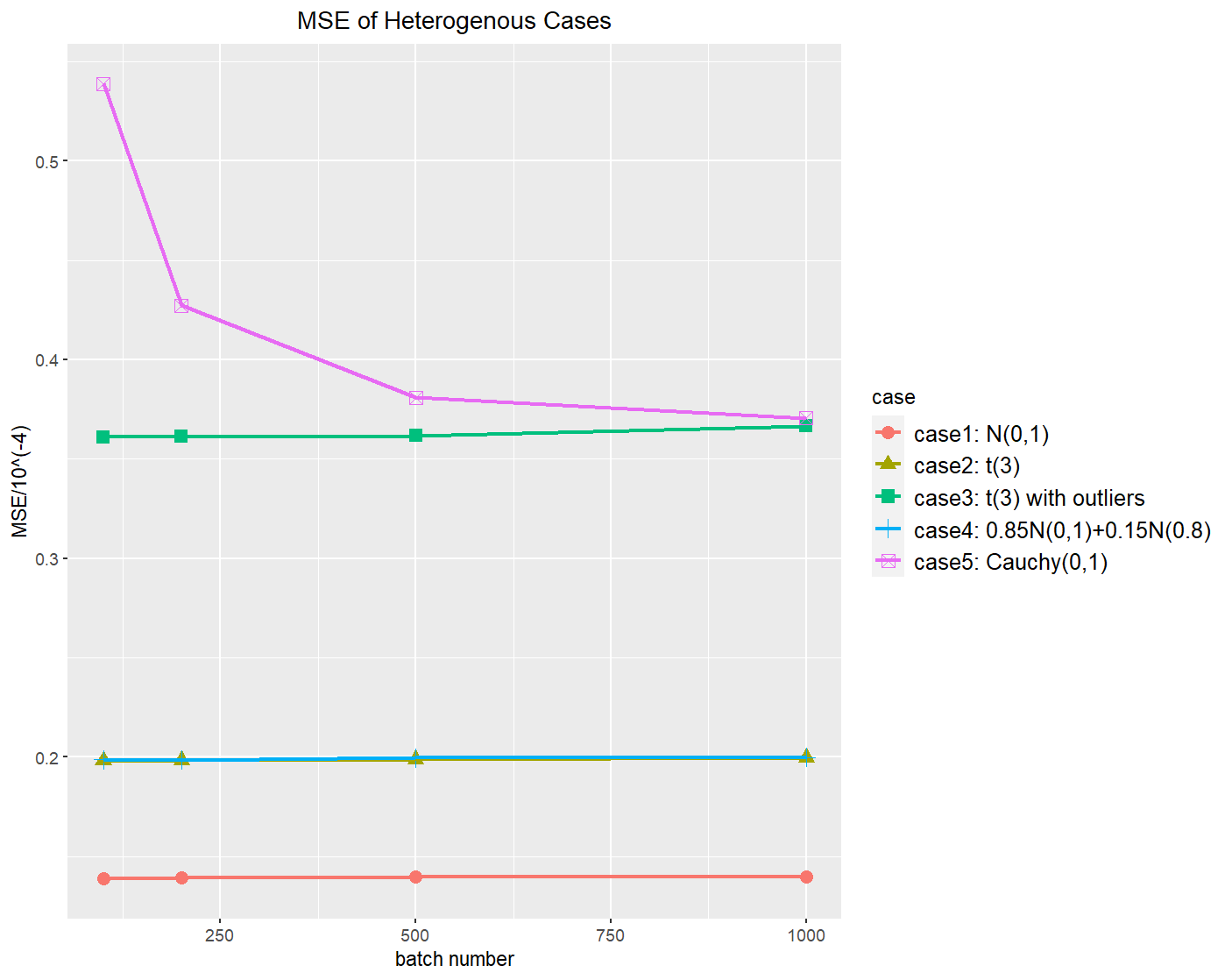}}
	\caption{MSE of UHR varies with batch number $b$}
	\label{graph1}
\end{figure}

(5) In order to better display the influence of subset number $b$ on estimating errors of UHR, we make a line graph showing the relationship between subset number $b$ and $MSE$ in heterogeneous cases(Figure \ref{graph1}). To sum up, for normal distribution, $t$ distribution, $t$ distribution with outliers and mixed symmetric normal distribution(case 1$\sim$case 4), estimation error only has subtle changes with $b$ increasing. That is, estimation results of UHR are robust to subset number. For Cauchy distribution case(case 5), estimation error goes down with $b$ increasing, as subset sample decreases, implying the same conclusion with (3).

\begin{table}[htbp]
	\centering
	\caption{Estimation Error in Experiment 1 for case 1}
	\begin{threeparttable}
			\begin{tabular}{c|c|cccc}
			\hline
			\multicolumn{6}{c}{case 1: N(0,1)} \\
			\hline
			&       & \multicolumn{4}{c}{homogeneous} \\
			\hline
			& $b$     & 100   & 200   & 500   & 1000 \\
			\hline
			\multirow{5}[2]{*}{$MSE$} & OLS   & \textbf{0.04122 } & \textbf{0.04122 } & \textbf{0.04122 } & \textbf{0.04122 } \\
			& RLS   & 0.04122  & 0.04122  & 0.04122  & 0.04122  \\
			& OHR   & 0.04325  & 0.04325  & 0.04325  & 0.04325  \\
			& DC-HR & 0.04335  & 0.04329  & 0.04336  & 0.04342  \\
			& UHR   & 0.04334  & 0.04333  & 0.04333  & 0.04336  \\
			\hline
			\multirow{5}[2]{*}{$MAE$} & OLS   & \textbf{0.32565 } & \textbf{0.32565 } & \textbf{0.32565 } & \textbf{0.32565 } \\
			& RLS   & 0.32565  & 0.32565  & 0.32565  & 0.32565  \\
			& OHR   & 0.33423  & 0.33423  & 0.33423  & 0.33423  \\
			& DC-HR & 0.33456  & 0.33414  & 0.33414  & 0.33500  \\
			& UHR   & 0.33458  & 0.33456  & 0.33454  & 0.33486  \\
			\hline
			&       & \multicolumn{4}{c}{heterogeneous} \\
			\hline
			& $b$     & 100   & 200   & 500   & 1000 \\
			\hline
			\multirow{5}[2]{*}{$MSE$} & OLS   & 0.24578  & 0.24578  & 0.24578  & 0.24578  \\
			& RLS   & 0.24578  & 0.24578  & 0.24578  & 0.24578  \\
			& OHR   & 0.13903  & \textbf{0.13903 } & \textbf{0.13903 } & \textbf{0.13903 } \\
			& DC-HR & \textbf{0.13879 } & 0.13911  & 0.13995  & 0.13974  \\
			& UHR   & 0.13884  & 0.13929  & 0.14001  & 0.13999  \\
			\hline
			\multirow{5}[2]{*}{$MAE$} & OLS   & 0.79640  & 0.79640  & 0.79640  & 0.79640  \\
			& RLS   & 0.79640  & 0.79640  & 0.79640  & 0.79640  \\
			& OHR   & 0.60189  & 0.60189  & \textbf{0.60189 } & \textbf{0.60189 } \\
			& DC-HR & \textbf{0.60066 } & \textbf{0.60161 } & 0.60297  & 0.60211  \\
			& UHR   & 0.60094  & 0.60205  & 0.60376  & 0.60315  \\
			\hline
		\end{tabular}%
	 \begin{tablenotes}    
		\footnotesize               
		\item[1] Units for $MSE$ and $MAE$ are $10^{-4}$ and  $10^{-2}$.          
		\item[2] The values shown in bold are the optimal indexes in each case.        
	\end{tablenotes}            
	\end{threeparttable}       
	\label{e1-case1}%
\end{table}%

\begin{table}[htbp]
	\centering
	\caption{Estimation Error in Experiment 1 for case 2}
	\begin{threeparttable}
		\begin{tabular}{c|c|cccc}
		\hline
		\multicolumn{6}{c}{case 2: $t(3)$} \\
		\hline
		&       & \multicolumn{4}{c}{homogeneous} \\
		\hline
		& $b$     & 100   & 200   & 500   & 1000 \\
		\hline
		\multirow{5}[2]{*}{$MSE$} & OLS   & 0.11885  & 0.11885  & 0.11885  & 0.11885  \\
		& RLS   & 0.11885  & 0.11885  & 0.11885  & 0.11885  \\
		& OHR   & 0.06468  & 0.06468  & \textbf{0.06468 } & \textbf{0.06468 } \\
		& DC-HR & 0.06479  & 0.06491  & 0.06500  & 0.06528  \\
		& UHR   & \textbf{0.06463 } & \textbf{0.06464 } & 0.06471  & 0.06473  \\
		\hline
		\multirow{5}[2]{*}{$MAE$} & OLS   & 0.54831  & 0.54831  & 0.54831  & 0.54831  \\
		& RLS   & 0.54831  & 0.54831  & 0.54831  & 0.54831  \\
		& OHR   & 0.40403  & 0.40403  & 0.40403  & 0.40403  \\
		& DC-HR & 0.40427  & 0.40454  & 0.40473  & 0.40503  \\
		& UHR   & \textbf{0.40381 } & \textbf{0.40383 } & \textbf{0.40394 } & \textbf{0.40364 } \\
		\hline
		&       & \multicolumn{4}{c}{heterogeneous} \\
		\hline
		& $b$     & 100   & 200   & 500   & 1000 \\
		\hline
		\multirow{5}[2]{*}{$MSE$} & OLS   & 0.74695  & 0.74695  & 0.74695  & 0.74695  \\
		& RLS   & 0.74695  & 0.74695  & 0.74695  & 0.74695  \\
		& OHR   & 0.19960  & 0.19960  & 0.19960  & 0.19960  \\
		& DC-HR & \textbf{0.19773 } & \textbf{0.19802 } & \textbf{0.19812 } & \textbf{0.19878 } \\
		& UHR   & 0.19823  & 0.19859  & 0.19902  & 0.19987  \\
		\hline
		\multirow{5}[2]{*}{$MAE$} & OLS   & 1.37955  & 1.37955  & 1.37955  & 1.37955  \\
		& RLS   & 1.37955  & 1.37955  & 1.37955  & 1.37955  \\
		& OHR   & 0.71031  & 0.71031  & 0.71031  & 0.71031  \\
		& DC-HR & \textbf{0.70748 } & \textbf{0.70796 } & \textbf{0.70763 } & \textbf{0.70854 } \\
		& UHR   & 0.70816  & 0.70844  & 0.70886  & 0.71044  \\
		\hline
	\end{tabular}%
	\begin{tablenotes}    
		\footnotesize               
		\item[1] Units for $MSE$ and $MAE$ are $10^{-4}$ and  $10^{-2}$.          
		\item[2] The values shown in bold are the optimal indexes in each case.        
	\end{tablenotes}            
	\end{threeparttable} 
	\label{e1-case2}%
\end{table}%

\begin{table}[htbp]
	\centering
	\caption{Estimation Error in Experiment 1 for case 3}
	\begin{threeparttable}
			\begin{tabular}{c|c|cccc}
			\hline
			\multicolumn{6}{c}{case 3: $t(3)$ with outliers} \\
			\hline
			&       & \multicolumn{4}{c}{homogeneous} \\
			\hline
			& $b$     & 100   & 200   & 500   & 1000 \\
			\hline
			\multirow{5}[2]{*}{$MSE$} & OLS   & 2.49242  & 2.49242  & 2.49242  & 2.49242  \\
			& RLS   & 2.49242  & 2.49242  & 2.49242  & 2.49242  \\
			& OHR   & 0.12271  & 0.12271  & 0.12271  & 0.12271  \\
			& DC-HR & 0.12207  & 0.12241  & 0.12333  & \textbf{0.12442 } \\
			& UHR   & \textbf{0.12185 } & \textbf{0.12199 } & \textbf{0.12258 } & 0.12310  \\
			\hline
			\multirow{5}[2]{*}{$MAE$} & OLS   & 2.54406  & 2.54406  & 2.54406  & 2.54406  \\
			& RLS   & 2.54406  & 2.54406  & 2.54406  & 2.54406  \\
			& OHR   & 0.55517  & 0.55517  & 0.55517  & \textbf{0.55517 } \\
			& DC-HR & 0.55402  & 0.55446  & 0.55679  & 0.55884  \\
			& UHR   & \textbf{0.55336 } & \textbf{0.55369 } & \textbf{0.55487 } & 0.55608  \\
			\hline
			&       & \multicolumn{4}{c}{heterogeneous} \\
			\hline
			& $b$     & 100   & 200   & 500   & 1000 \\
			\hline
			\multirow{5}[2]{*}{$MSE$} & OLS   & 3.18120  & 3.18120  & 3.18120  & 3.18120  \\
			& RLS   & 3.18120  & 3.18120  & 3.18120  & 3.18120  \\
			& OHR   & 0.36930  & 0.36930  & 0.36930  & 0.36930  \\
			& DC-HR & 0.36105  & 0.36105  & 0.36166  & 0.36678  \\
			& UHR   & \textbf{0.36110 } & \textbf{0.36135 } & \textbf{0.36180 } & \textbf{0.36654 } \\
			\hline
			\multirow{5}[2]{*}{$MAE$} & OLS   & 2.87060  & 2.87060  & 2.87060  & 2.87060  \\
			& RLS   & 2.87060  & 2.87060  & 2.87060  & 2.87060  \\
			& OHR   & 0.96724  & 0.96724  & 0.96724  & 0.96724  \\
			& DC-HR & 0.95639  & 0.95577  & 0.95761  & 0.96439  \\
			& UHR   & \textbf{0.95605 } & \textbf{0.95594 } & \textbf{0.95672 } & \textbf{0.96274 } \\
			\hline
		\end{tabular}%
	\begin{tablenotes}    
		\footnotesize               
		\item[1] Units for $MSE$ and $MAE$ are $10^{-4}$ and  $10^{-2}$.          
		\item[2] The values shown in bold are the optimal indexes in each case.        
	\end{tablenotes}            
	\end{threeparttable} 
	\label{e1-case3}%
\end{table}%

\begin{table}[htbp]
	\centering
	\caption{Estimation Error in experiment 1 for case 4}
	\begin{threeparttable}
			\begin{tabular}{c|c|cccc}
			\hline
			\multicolumn{6}{c}{case 4: 0.85×N(0,1)+0.15×N(0,8)} \\
			\hline
			&       & \multicolumn{4}{c}{homogeneous} \\
			\hline
			& $b$     & 100   & 200   & 500   & 1000 \\
			\hline
			\multirow{5}[2]{*}{$MSE$} & OLS   & 0.07771  & 0.07771  & 0.07771  & 0.07771  \\
			& RLS   & 0.07771  & 0.07771  & 0.07771  & 0.47451  \\
			& OHR   & 0.05530  & 0.05530  & 0.05530  & 0.05530  \\
			& DC-HR & 0.05538  & 0.05543  & 0.05544  & 0.05565  \\
			& UHR   & \textbf{0.05524 } & \textbf{0.05524 } & \textbf{0.05518 } & \textbf{0.05519 } \\
			\hline
			\multirow{5}[2]{*}{$MAE$} & OLS   & 0.44224  & 0.44224  & 0.44224  & 0.44224  \\
			& RLS   & 0.44224  & 0.44224  & 0.44224  & 1.11088  \\
			& OHR   & 0.37259  & 0.37259  & 0.37259  & 0.37259  \\
			& DC-HR & 0.37309  & 0.37324  & 0.37331  & 0.37356  \\
			& UHR   & \textbf{0.37238 } & \textbf{0.37240 } & \textbf{0.37233 } & \textbf{0.37205 } \\
			\hline
			&       & \multicolumn{4}{c}{heterogeneous} \\
			\hline
			& $b$     & 100   & 200   & 500   & 1000 \\
			\hline
			\multirow{5}[2]{*}{$MSE$} & OLS   & 0.47451  & 0.47451  & 0.47451  & 0.47451  \\
			& RLS   & 0.47451  & 0.47451  & 0.47451  & 0.47451  \\
			& OHR   & 0.19964  & 0.19964  & 0.19964  & 0.19964  \\
			& DC-HR & 0.19859  & 0.19857  & \textbf{0.19953 } & \textbf{0.19965 } \\
			& UHR   & \textbf{0.19881 } & \textbf{0.19866 } & 0.19967  & 0.19990  \\
			\hline
			\multirow{5}[2]{*}{$MAE$} & OLS   & 1.11088  & 1.11088  & 1.11088  & 1.11088  \\
			& RLS   & 1.11088  & 1.11088  & 1.11088  & 1.11088  \\
			& OHR   & 0.71519  & 0.71519  & 0.71519  & 0.71519  \\
			& DC-HR & \textbf{0.71321 } & \textbf{0.71335 } & \textbf{0.71535 } & \textbf{0.71578 } \\
			& UHR   & 0.71364  & 0.71377  & 0.71559  & 0.71599  \\
			\hline
		\end{tabular}%
		\begin{tablenotes}    
		\footnotesize               
		\item[1] Units for $MSE$ and $MAE$ are $10^{-4}$ and  $10^{-2}$.          
		\item[2] The values shown in bold are the optimal indexes in each case.        
		\end{tablenotes}            
	\end{threeparttable} 
	\label{e1-case4}%
\end{table}%

\begin{table}[htbp]
	\centering
	\caption{Estimation Error in Experiment 1 for case 5}
	\begin{threeparttable}
			\begin{tabular}{c|c|cccc}
			\hline
			\multicolumn{6}{c}{case 5: Cauchy(0,1)} \\
			\hline
			&       & \multicolumn{4}{c}{homogeneous} \\
			\hline
			& $b$     & 100   & 200   & 500   & 1000 \\
			\hline
			\multirow{5}[2]{*}{$MSE$} & OLS   & 11711466.60880  & 11711466.60880  & 11711466.60880  & 11711466.60880  \\
			& RLS   & 11711466.60880  & 11711466.60880  & 11711466.60880  & 11711466.60880  \\
			& OHR   & 7.16525  & 7.16525  & 7.16525  & 7.16525  \\
			& DC-HR & 0.15638  & 0.14910  & 0.14287  & 0.14240  \\
			& UHR   & \textbf{0.15626 } & \textbf{0.14878 } & \textbf{0.14190 } & \textbf{0.14109 } \\
			\hline
			\multirow{5}[2]{*}{$MAE$} & OLS   & 1502.46026  & 1502.46026  & 1502.46026  & 1502.46026  \\
			& RLS   & 1502.46026  & 1502.46026  & 1502.46026  & 1502.46026  \\
			& OHR   & 3.23099  & 3.23099  & 3.23099  & 3.23099  \\
			& DC-HR & 0.63014  & 0.61556  & 0.60233  & 0.60258  \\
			& UHR   & \textbf{0.63009 } & \textbf{0.61492 } & \textbf{0.60000 } & \textbf{0.60016 } \\
			\hline
			&       & \multicolumn{4}{c}{heterogeneous} \\
			\hline
			& b     & 100   & 200   & 500   & 1000 \\
			\hline
			\multirow{5}[2]{*}{$MSE$} & OLS   & 29518873.58901  & 29518873.58901  & 29518873.58901  & 29518873.58901  \\
			& RLS   & 29518873.58901  & 29518873.58901  & 29518873.58901  & 29518873.58901  \\
			& OHR   & 102.19791  & 102.19791  & 102.19791  & 102.19791  \\
			& DC-HR & \textbf{0.53841 } & \textbf{0.42683 } & 0.38114  & 0.37118  \\
			& UHR   & 0.53882  & 0.42715  & \textbf{0.38089 } & \textbf{0.37050 } \\
			\hline
			\multirow{5}[2]{*}{$MAE$} & OLS   & 2651.78745  & 2651.78745  & 2651.78745  & 2651.78745  \\
			& RLS   & 2651.78745  & 2651.78745  & 2651.78745  & 2651.78745  \\
			& OHR   & 13.48746  & 13.48746  & 13.48746  & 13.48746  \\
			& DC-HR & \textbf{1.16842 } & \textbf{1.04156 } & 0.98940  & 0.97666  \\
			& UHR   & 1.16899  & 1.04247  & \textbf{0.98857 } & \textbf{0.97579 } \\
			\hline
		\end{tabular}%
		\begin{tablenotes}    
		\footnotesize               
		\item[1] Units for $MSE$ and $MAE$ are $10^{-4}$ and  $10^{-2}$.          
		\item[2] The values shown in bold are the optimal indexes in each case.        
	\end{tablenotes}            
	\end{threeparttable} 
	\label{e1-case6}%
\end{table}%

Table \ref{e2-case1} to Table \ref{e2-case6} show the results of \textbf{Experiment\ 2}, where $b$ varies from $10$ to $10000$ while subset number $n_t$ keeps fixed. We can obtain following conclusions through these tables.

(1) With $b$ increasing, all the algorithms are performing better. However, when $N_b$ is small(for example, $N_b=1000$) and random error is heavily contaminated(such as case 4,5 with heterogeneous error), estimation error of OLS is much larger compared with Huber algorithms. UHR also performs rather well, which indicates our algorithm suits for not only big data streams but also normal-scale datasets.

(2) For case 1$\sim$case 4, OHR has the smallest estimation error. Although UHR does not perform better than OHR, it only has subtle differences with OHR even if $N_b$ is in a small amount, indicating that UHR is able to achieve good asymptotic equivalency with OHR.

(3) For case 5, the estimation error of UHR is much smaller than that of OHR. This points to the same conclusion with \textbf{Experiment 1}, that is, smaller subsets may have better estimation efficiency than entire dataset for Cauchy distribution.

(4) UHR algorithm performs better in most cases than DC-HR, proving the superiority of online updating algorithm.

(5) At last, in settings, \textbf{Experiment 2} is close to big data streams. Each subset has a small size with 100 samples, while subset number $b$ increases from 10 to 10000. This is just like the process of new subsets arriving in big data streams. We can find that the estimation error of UHR decreases with subset number increasing and is similar to OHR, which implies UHR is applicable to big data streams.

\begin{table}[htbp]
	\centering
	\caption{Estimation Error in Experiment 2 for case 1}
	\begin{threeparttable} 
	\begin{tabular}{c|c|cccc}
		\hline
		\multicolumn{6}{c}{case 1: N(0,1)} \\
		\hline
		&       & \multicolumn{4}{c}{homogeneous} \\
		\hline
		& $N$ & $10^3$ & $10^4$ & $10^5$ & $10^6$ \\
		\hline
		\multirow{5}[2]{*}{$MSE$} & OLS   & \textbf{41.65896 } & \textbf{3.88942 } & \textbf{0.39393 } & \textbf{0.04122 } \\
		& RLS   & 41.65896  & 3.88942  & 0.39393  & 0.04122  \\
		& OHR   & 44.27814  & 4.07750  & 0.41684  & 0.04325  \\
		& DC-HR & 46.34740  & 4.35515  & 0.44800  & 0.04593  \\
		& UHR   & 44.27868  & 4.09200  & 0.42152  & 0.04357  \\
		\hline
		\multirow{5}[2]{*}{$MAE$} & OLS   & \textbf{10.40966 } & \textbf{3.18110 } & \textbf{1.00481 } & \textbf{0.32565 } \\
		& RLS   & 10.40966  & 3.18110  & 1.00481  & 0.32565  \\
		& OHR   & 10.74373  & 3.25098  & 1.03314  & 0.33423  \\
		& DC-HR & 10.94708  & 3.35412  & 1.06948  & 0.34447  \\
		& UHR   & 10.7334727* & 3.2551126* & 1.0428132* & 0.3357810* \\
		\hline
		&       & \multicolumn{4}{c}{heterogeneous} \\
		\hline
		& $N$ & $10^3$ & $10^4$ & $10^5$ & $10^6$ \\
		\hline
		\multirow{5}[2]{*}{$MSE$} & OLS   & 227.80014  & 23.45107  & 2.41744  & 0.24578  \\
		& RLS   & 227.80014  & 23.45107  & 2.41744  & 0.24578  \\
		& OHR   & \textbf{146.58109 } & \textbf{14.57030 } & \textbf{1.38268 } & \textbf{0.13903 } \\
		& DC-HR & 153.32625* & 15.35639* & 1.44728  & 0.14974  \\
		& UHR   & 153.78150  & 15.44100  & 1.44311* & 0.14645* \\
		\hline
		\multirow{5}[2]{*}{$MAE$} & OLS   & 24.24029  & 7.76669  & 2.49384  & 0.79640  \\
		& RLS   & 24.24029  & 7.76669  & 2.49384  & 0.79640  \\
		& OHR   & \textbf{19.38650 } & \textbf{6.11912 } & \textbf{1.87599 } & \textbf{0.60189 } \\
		& DC-HR & 19.80401* & 6.27307* & 1.91918  & 0.61996  \\
		& UHR   & 19.84922  & 6.31367  & 1.91440* & 0.61727* \\
		\hline
	\end{tabular}%
		\begin{tablenotes}    
		\footnotesize               
		\item[1] Units for $MSE$ and $MAE$ are $10^{-4}$ and  $10^{-2}$.          
		\item[2] The values shown in bold are the optimal indexes in each case.        
		\item[3] * represents the superior of UHR and DC-HR.
	\end{tablenotes}            
	\end{threeparttable} 
	\label{e2-case1}%
\end{table}%

\begin{table}[htbp]
	\centering
	\caption{Estimation Error in Experiment 2 for case 2}
	\begin{threeparttable} 
			\begin{tabular}{c|c|cccc}
			\hline
			\multicolumn{6}{c}{case 2: $t(3)$} \\
			\hline
			&       & \multicolumn{4}{c}{homogeneous} \\
			\hline
			& $N$ & $10^3$ & $10^4$ & $10^5$ & $10^6$ \\
			\hline
			\multirow{5}[2]{*}{$MSE$} & OLS   & 117.55450  & 11.13197  & 1.16896  & 0.11885  \\
			& RLS   & 117.55450  & 11.13197  & 1.16896  & 0.11885  \\
			& OHR   & \textbf{65.99914 } & \textbf{6.21516 } & \textbf{0.61343 } & \textbf{0.06468 } \\
			& DC-HR & 70.83631  & 6.79250  & 0.66002  & 0.07046  \\
			& UHR   & 67.51753  & 6.35419  & 0.62871  & 0.06651  \\
			\hline
			\multirow{5}[2]{*}{$MAE$} & OLS   & 17.16898  & 5.29056  & 1.73030  & 0.54831  \\
			& RLS   & 17.16898  & 5.29056  & 1.73030  & 0.54831  \\
			& OHR   & \textbf{13.04886 } & \textbf{3.96196 } & \textbf{1.25701 } & \textbf{0.40403 } \\
			& DC-HR & 13.47303  & 4.12096  & 1.30451  & 0.41925  \\
			& UHR   & 13.16859* & 4.00056* & 1.27431* & 0.40905* \\
			\hline
			&       & \multicolumn{4}{c}{heterogeneous} \\
			\hline
			& $N$ & $10^3$ & $10^4$ & $10^5$ & $10^6$ \\
			\hline
			\multirow{5}[2]{*}{$MSE$} & OLS   & 713.28555  & 72.67236  & 7.44372  & 0.74695  \\
			& RLS   & 713.28555  & 72.67236  & 7.44372  & 0.74695  \\
			& OHR   & \textbf{210.04680 } & \textbf{21.71120 } & \textbf{1.38268 } & \textbf{0.19960 } \\
			& DC-HR & 223.54581* & 23.37703  & 2.21219* & 0.21934  \\
			& UHR   & 225.35099  & 23.18896* & 2.23103  & 0.21598* \\
			\hline
			\multirow{5}[2]{*}{$MAE$} & OLS   & 41.74305  & 13.58001  & 4.35484  & 1.37955  \\
			& RLS   & 41.74305  & 13.58001  & 4.35484  & 1.37955  \\
			& OHR   & \textbf{22.96835 } & \textbf{7.49359 } & \textbf{1.87599 } & \textbf{0.71031 } \\
			& DC-HR & 23.71750* & 7.79915  & 2.38534* & 0.74458  \\
			& UHR   & 23.79698  & 7.72857* & 2.39061  & 0.73697* \\
			\hline
		\end{tabular}%
		\begin{tablenotes}    
		\footnotesize               
		\item[1] Units for $MSE$ and $MAE$ are $10^{-4}$ and  $10^{-2}$.          
		\item[2] The values shown in bold are the optimal indexes in each case.        
		\item[3] * represents the superior of UHR and DC-HR.
	\end{tablenotes}            
	\end{threeparttable}
	\label{e2-case2}%
\end{table}%

\begin{table}[htbp]
	\centering
	\caption{Estimation Error in Experiment 2 for case 3}
	\begin{threeparttable}
		\begin{tabular}{c|c|cccc}
		\hline
		\multicolumn{6}{c}{case 3: $t(3)$ with outliers} \\
		\hline
		&       & \multicolumn{4}{c}{homogeneous} \\
		\hline
		& $N$ & $10^3$ & $10^4$ & $10^5$ & $10^6$ \\
		\hline
		\multirow{5}[2]{*}{$MSE$} & OLS   & 2507.30950  & 246.33631  & 25.46565  & 2.49242  \\
		& RLS   & 2507.30950  & 246.33631  & 25.46565  & 2.49242  \\
		& OHR   & \textbf{122.73570 } & \textbf{11.76220 } & \textbf{1.12463 } & \textbf{0.12271 } \\
		& DC-HR & 147.05309  & 14.42111  & 1.37263  & 0.15004  \\
		& UHR   & 140.65747  & 13.72919  & 1.30735  & 0.14124  \\
		\hline
		\multirow{5}[2]{*}{$MAE$} & OLS   & 79.30791  & 25.08737  & 8.04774  & 2.54406  \\
		& RLS   & 79.30791  & 25.08737  & 8.04774  & 2.54406  \\
		& OHR   & \textbf{17.63695 } & \textbf{5.45740 } & \textbf{1.67765 } & \textbf{0.55517 } \\
		& DC-HR & 19.35555  & 6.00961  & 1.86798  & 0.61847  \\
		& UHR   & 18.83415* & 5.84983* & 1.81094* & 0.59913* \\
		\hline
		&       & \multicolumn{4}{c}{heterogeneous} \\
		\hline
		& $N$ & $10^3$ & $10^4$ & $10^5$ & $10^6$ \\
		\hline
		\multirow{5}[2]{*}{$MSE$} & OLS   & 3160.12456  & 304.54568  & 30.63899  & 3.18120  \\
		& RLS   & 3160.12456  & 304.54568  & 30.63899  & 3.18120  \\
		& OHR   & \textbf{378.19492 } & \textbf{37.93189 } & \textbf{3.64422 } & \textbf{0.36930 } \\
		& DC-HR & 449.14772  & 45.55376  & 4.46745  & 0.44603  \\
		& UHR   & 443.33399* & 44.43532* & 4.36829* & 0.43609* \\
		\hline
		\multirow{5}[2]{*}{$MAE$} & OLS   & 89.66953  & 28.08533  & 8.77479  & 2.87060  \\
		& RLS   & 89.66953  & 28.08533  & 8.77479  & 2.87060  \\
		& OHR   & \textbf{30.89182 } & \textbf{9.82228 } & \textbf{3.03822 } & \textbf{0.96724 } \\
		& DC-HR & 33.87220  & 10.77172  & 3.37132  & 1.06330  \\
		& UHR   & 33.60676* & 10.62887* & 3.34340* & 1.05017* \\
		\hline
	\end{tabular}%
		\begin{tablenotes}    
		\footnotesize               
		\item[1] Units for $MSE$ and $MAE$ are $10^{-4}$ and  $10^{-2}$.          
		\item[2] The values shown in bold are the optimal indexes in each case.        
		\item[3] * represents the superior of UHR and DC-HR.
	\end{tablenotes}            
	\end{threeparttable} 
	\label{e2-case3}%
\end{table}%

\begin{table}[htbp]
	\centering
	\caption{Estimation Error in Experiment 2 for case 4}
	\begin{threeparttable}
			\begin{tabular}{c|c|cccc}
			\hline
			\multicolumn{6}{c}{case 4: 0.85×N(0,1)+0.15×N(0,8)} \\
			\hline
			&       & \multicolumn{4}{c}{homogeneous} \\
			\hline
			& $N$ & $10^3$ & $10^4$ & $10^5$ & $10^6$ \\
			\hline
			\multirow{5}[2]{*}{$MSE$} & OLS   & 83.55944  & 8.00806  & 0.77406  & 0.07771  \\
			& RLS   & 83.55944  & 8.00806  & 0.77406  & 0.07771  \\
			& OHR   & \textbf{56.64867 } & \textbf{5.47834 } & \textbf{0.55143 } & \textbf{0.05530 } \\
			& DC-HR & 59.51882  & 5.80763  & 0.58022  & 0.05906  \\
			& UHR   & 57.78717* & 5.56164* & 0.56295* & 0.05617* \\
			\hline
			\multirow{5}[2]{*}{$MAE$} & OLS   & 14.42020  & 4.52082  & 1.38455  & 0.44224  \\
			& RLS   & 14.42020  & 4.52082  & 1.38455  & 0.44224  \\
			& OHR   & \textbf{11.89712 } & \textbf{3.78035 } & \textbf{1.17328 } & \textbf{0.37259 } \\
			& DC-HR & 12.18530  & 3.89420  & 1.19697  & 0.38471  \\
			& UHR   & 11.97547* & 3.82015* & 1.18172* & 0.37528* \\
			\hline
			&       & \multicolumn{4}{c}{heterogeneous} \\
			\hline
			& $N$ & $10^3$ & $10^4$ & $10^5$ & $10^6$ \\
			\hline
			\multirow{5}[2]{*}{$MSE$} & OLS   & 472.41924  & 51.87671  & 14382.78626  & 0.47451  \\
			& RLS   & 472.41924  & 51.87671  & 14382.78626  & 0.47451  \\
			& OHR   & \textbf{194.14143 } & \textbf{19.97540 } & \textbf{1.94836 } & \textbf{0.19964 } \\
			& DC-HR & 208.04891  & 21.31645  & 2.04540* & 0.20936* \\
			& UHR   & 206.90744* & 21.03077* & 2.06516  & 0.21083  \\
			\hline
			\multirow{5}[2]{*}{$MAE$} & OLS   & 34.72670  & 11.48330  & 239.77327  & 1.11088  \\
			& RLS   & 34.72670  & 11.48330  & 239.77327  & 1.11088  \\
			& OHR   & \textbf{22.14798 } & \textbf{7.18230 } & \textbf{2.24135 } & \textbf{0.71519 } \\
			& DC-HR & 22.89192  & 7.35959  & 2.30214* & 0.73328* \\
			& UHR   & 22.88671* & 7.30928* & 2.30965  & 0.73398  \\
			\hline
		\end{tabular}%
	\begin{tablenotes}    
		\footnotesize               
		\item[1] Units for $MSE$ and $MAE$ are $10^{-4}$ and  $10^{-2}$.          
		\item[2] The values shown in bold are the optimal indexes in each case.        
		\item[3] * represents the superior of UHR and DC-HR.
	\end{tablenotes}            
	\end{threeparttable} 
	\label{e2-case4}%
\end{table}%

\begin{table}[htbp]
	\centering
	\caption{Estimation Error in Experiment 2 for case 5}
	\begin{threeparttable}
			\begin{tabular}{c|c|cccc}
			\hline
			\multicolumn{6}{c}{case 5: Cauchy(0,1)} \\
			\hline
			&       & \multicolumn{4}{c}{homogeneous} \\
			\hline
			& $N$ & $10^3$ & $10^4$ & $10^5$ & $10^6$ \\
			\hline
			\multirow{5}[2]{*}{$MSE$} & OLS   & 13031861.67103  & 37026406.50190  & 6041016.96141  & 11711466.60880  \\
			& RLS   & 13031861.67103  & 37026406.50190  & 6041016.96141  & 11711466.60880  \\
			& OHR   & \textbf{146.07957 } & 16.91701  & 3.81870  & 7.16525  \\
			& DC-HR & 169.68684  & 17.13059  & 1.69887  & 0.16616  \\
			& UHR   & 162.31759* & \textbf{16.29104*} & \textbf{1.61969*} & \textbf{0.15643*} \\
			\hline
			\multirow{5}[2]{*}{$MAE$} & OLS   & 1431.93975  & 1881.11733  & 1147.11445  & 1502.46026  \\
			& RLS   & 1431.93975  & 1881.11733  & 1147.11445  & 1502.46026  \\
			& OHR   & \textbf{19.26524 } & \textbf{6.32966 } & 2.65689  & 3.23099  \\
			& DC-HR & 20.69384  & 6.62144  & 2.08405  & 0.65167  \\
			& UHR   & 20.25285* & 6.45609* & \textbf{2.03275*} & \textbf{0.63216*} \\
			\hline
			&       & \multicolumn{4}{c}{heterogeneous} \\
			\hline
			& $N$ & $10^3$ & $10^4$ & $10^5$ & $10^6$ \\
			\hline
			\multirow{5}[2]{*}{$MSE$} & OLS   & 47087625.58229  & 123184496.07168  & 92413723.44581  & 29518873.58901  \\
			& RLS   & 47087625.58229  & 123184496.07168  & 92413723.44581  & 29518873.58901  \\
			& OHR   & \textbf{362.02026 } & 3340.24291  & 40.58347  & 102.19791  \\
			& DC-HR & 418.52258  & 46.02272  & 4.17859  & 0.42415  \\
			& UHR   & 413.20117* & \textbf{45.54626*} & \textbf{4.13358*} & \textbf{0.41610*} \\
			\hline
			\multirow{5}[2]{*}{$MAE$} & OLS   & 2705.91096  & 3340.24291  & 2526.77046  & 2651.78745  \\
			& RLS   & 2705.91096  & 3340.24291  & 2526.77046  & 2651.78745  \\
			& OHR   & \textbf{30.23063 } & 11.18484  & 7.40727  & 13.48746  \\
			& DC-HR & 32.69547  & 10.84272  & 3.26736  & 1.05280  \\
			& UHR   & 32.45720* & \textbf{10.82107*} & \textbf{3.23621*} & \textbf{1.03787*} \\
			\hline
		\end{tabular}%
		\begin{tablenotes}    
		\footnotesize               
		\item[1] Units for $MSE$ and $MAE$ are $10^{-4}$ and  $10^{-2}$.          
		\item[2] The values shown in bold are the optimal indexes in each case.        
		\item[3] * represents the superior of UHR and DC-HR.
	\end{tablenotes}            
	\end{threeparttable} 
	\label{e2-case6}%
\end{table}%

\subsubsection{Calculation Efficiency}
In calculation efficiency, we mainly focus on \textbf{Experiment\ 1}, \textbf{Experiment\ 3} and \textbf{Experiment\ 4}. \textbf{Experiment\ 1} investigates the influence of $b$ on calculation efficiency when total sample size $N_b$ is fixed. \textbf{Experiment\ 3} focuses on the influence of $N_b$ when $n_t$ is fixed and relatively large. \textbf{Experiment\ 4} is rather similar to \textbf{Experiment\ 3}, whereas, its fixed $n_t$ is smaller than \textbf{Experiment\ 3}, aiming at comparing the differences between different scales of $n_t$.

\begin{figure}
	\centerline{\includegraphics[height=6.5cm]{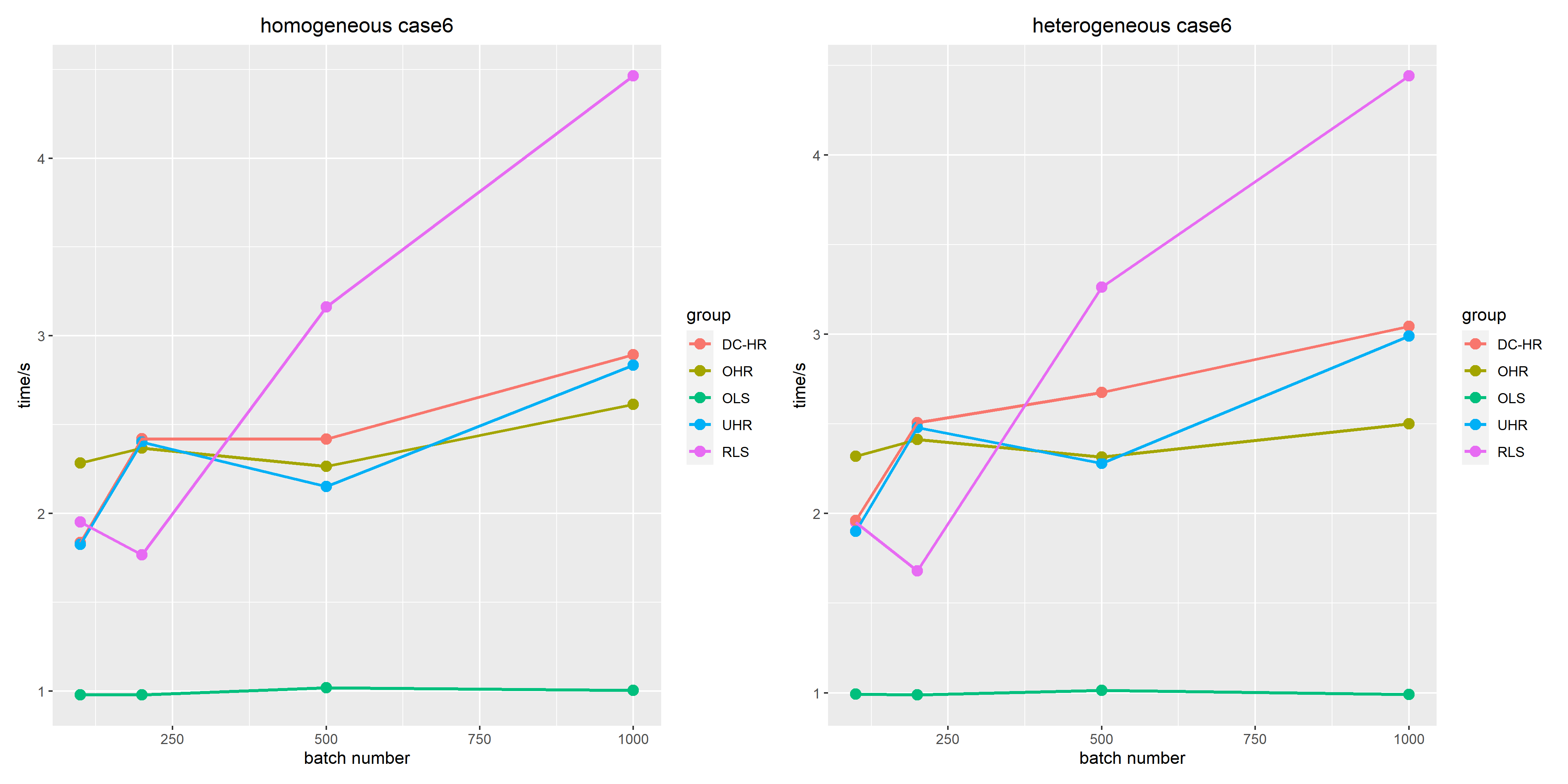}}
	\caption{batch number-calculation time for case6}
	\label{graph2}
\end{figure}

For \textbf{Experiment\ 1}, we can obtain the average running time of certain random error according to Eq.(\ref{time}). Before analyzing, it should be mentioned that OLS and OHR, which use entire dataset, should have had no difference in running time due to fixed $N_b$. However, in execution, we run OLS and OHR 4 times in different $b$ value settings. This may result in random effect on computer, leading to small differences in running time of OLS and RLS. Details are shown in Table \ref{e3-b100} to Table \ref{e3-b1000}. Based on these results, we can make following conclusions:

(1) Our UHR runs faster than OHR, DC-HR and RLS in most cases. With $b$ increasing, running time of RLS jumps up. Although running time of UHR increases as well, the trend is more slowly than RLS, indicating better property of UHR. It can be seen in Figure \ref{graph2} for details. As the results of 6 cases are similar, we only visualize case 6.

(2)Overall, Linear Regression algorithms perform better in calculation efficiency compared with Huber algorithms. However, they are not robust to contaminated random error according to \textbf{Experiment\ 1}. As a result, we still prefer UHR algorithm to Linear Regression.

(3)When subset number $b$ is rather small(like $b=100, 200, 500$), UHR has better calculation efficiency than OHR. However, this may not come true when $b$ increases to 1000. A possible reason is that with $b$ increasing, iteration increases as well. Accumulated running time due to iterations leads to a longer time. To solve it, we can artificially set $b$ in a smaller value for $b$ has no effect on estimation efficiency. In big data streams, we tend to prefer a rather small subset size $n_t$ then setting small $b$ for higher calculation efficiency may not work. However, as entire dataset cannot be obtained in big data streams scenario, OHR is no longer applicable and UHR is the only possible method so that a little bit longer running time of UHR is tolerable.

\begin{table}[htbp]
	\centering
	\caption{Average Running Time in Experiment 3 for $b=100$}
	\begin{threeparttable}
	\begin{tabular}{c|cccccc}
		\hline
		\multicolumn{7}{c}{$b=100$} \\
		\hline
		& case 1 & case 2 & case 3 & case 4 & case 5 & case 6 \\
		\hline
		& \multicolumn{6}{c}{homogeneous} \\
		\hline
		OLS   & 0.98250  & 0.98310  & 0.95790  & 0.98050  & 0.97960  & 0.97840  \\
		RLS   & 1.96290  & 1.94380  & 1.95850  & 1.97450  & 1.96220  & 1.95190  \\
		OHR   & 2.02670  & 2.28570  & 2.37480  & 2.22820  & 2.29830  & 2.28450  \\
		DC-HR & 1.77900  & 1.86630  & 1.87670  & 1.83970  & 1.77620  & 1.83630  \\
		UHR   & \textbf{1.75810 } & \textbf{1.84130 } & \textbf{1.83750 } & \textbf{1.81400 } & \textbf{1.85240 } & \textbf{1.82630 } \\
		\hline
		& \multicolumn{6}{c}{heterogeneous} \\
		\hline
		OLS   & 0.97090  & 0.97520  & 0.97280  & 0.97380  & 0.97270  & 0.99380  \\
		RLS   & \textbf{1.96890 } & 2.04620  & 1.97240  & \textbf{1.95330 } & \textbf{1.96670 } & 1.95050  \\
		OHR   & 2.52180  & 2.57020  & 2.45510  & 2.52730  & 4.55150  & 2.32000  \\
		DC-HR & 2.02770  & \textbf{1.97980 } & 2.00170  & 2.07250  & 2.49510  & 1.96030  \\
		UHR   & 2.01890  & 2.03020  & \textbf{1.93080 } & 2.04700  & 2.46830  & \textbf{1.90100 } \\
		\hline
	\end{tabular}%
	\begin{tablenotes}    
		\footnotesize               
		\item[1] The values shown in bold are the optimal model indexes (except OLS algorithm) in each case.        
	\end{tablenotes}            
	\end{threeparttable} 
	\label{e3-b100}%
\end{table}%

\begin{table}[htbp]
	\centering
	\caption{Average Running Time in Experiment 3 for $b=200$}
	\begin{threeparttable}
			\begin{tabular}{c|cccccc}
			\hline
			\multicolumn{7}{c}{$b=200$} \\
			\hline
			& case 1 & case 2 & case 3 & case 4 & case 5 & case 6 \\
			\hline
			& \multicolumn{6}{c}{homogeneous} \\
			\hline
			OLS   & 0.97720  & 0.98290  & 1.01140  & 0.98460  & 0.98500  & 0.97810  \\
			RLS   & \textbf{1.67800 } & \textbf{1.65770 } & \textbf{1.65050 } & \textbf{1.65270 } & \textbf{1.64270 } & \textbf{1.76660 } \\
			OHR   & 2.10430  & 2.40010  & 2.46390  & 2.32010  & 2.42110  & 2.36800  \\
			DC-HR & 2.36680  & 2.43230  & 2.42160  & 2.41890  & 2.42460  & 2.42040  \\
			UHR   & 2.34550  & 2.40360  & 2.38780  & 2.39220  & 2.37850  & 2.40130  \\
			\hline
			& \multicolumn{6}{c}{heterogeneous} \\
			\hline
			OLS   & 1.02260  & 0.98390  & 0.97880  & 1.01350  & 0.97970  & 0.98880  \\
			RLS   & \textbf{1.75910 } & \textbf{1.61520 } & \textbf{1.77020 } & \textbf{1.69830 } & \textbf{1.61170 } & \textbf{1.67900 } \\
			OHR   & 2.61610  & 2.68210  & 2.54390  & 2.64820  & 4.81750  & 2.41370  \\
			DC-HR & 2.62560  & 2.63130  & 2.53630  & 2.64060  & 2.97750  & 2.50600  \\
			UHR   & 2.59650  & 2.57810  & 2.50440  & 2.59220  & 2.91410  & 2.47980  \\
			\hline
		\end{tabular}%
	\begin{tablenotes}    
		\footnotesize               
		\item[1] The values shown in bold are the optimal model indexes (except OLS algorithm) in each case.        
	\end{tablenotes}            
	\end{threeparttable} 
	\label{e3-b200}%
\end{table}%

\begin{table}[htbp]
	\centering
	\caption{Average Running Time in Experiment 3 for $b=500$}
	\begin{threeparttable}
			\begin{tabular}{c|cccccc}
			\hline
			\multicolumn{7}{c}{$b=500$} \\
			\hline
			& case 1 & case 2 & case 3 & case 4 & case 5 & case 6 \\
			\hline
			& \multicolumn{6}{c}{homogeneous} \\
			\hline
			OLS   & 1.00940  & 0.99870  & 1.01550  & 1.02750  & 1.08760  & 1.01830  \\
			RLS   & 3.42860  & 3.29650  & 3.14050  & 3.29470  & 3.16500  & 3.16210  \\
			OHR   & \textbf{2.01170 } & 2.39080  & 2.30080  & 2.33080  & 2.35250  & 2.26460  \\
			DC-HR & 2.34690  & 2.53610  & 2.38140  & 2.44700  & 2.52070  & 2.41760  \\
			UHR   & 2.11160  & \textbf{2.19880 } & \textbf{2.17990 } & \textbf{2.13890 } & \textbf{2.19090 } & \textbf{2.15200 } \\
			\hline
			& \multicolumn{6}{c}{heterogeneous} \\
			\hline
			OLS   & 1.02390  & \textbf{0.98440 } & 1.09730  & 1.01810  & 1.01980  & 1.01280  \\
			RLS   & 3.54000  & 3.19630  & 3.15630  & 3.24660  & 3.16610  & 3.26170  \\
			OHR   & 2.47570  & 2.59900  & 2.36000  & 2.62570  & 4.64180  & 2.31450  \\
			DC-HR & 2.71570  & 2.92090  & 2.62800  & 2.97190  & 3.29770  & 2.67450  \\
			UHR   & \textbf{2.50870 } & 2.66040  & \textbf{2.35530 } & \textbf{2.59340 } & \textbf{2.87450 } & \textbf{2.27890 } \\
			\hline
		\end{tabular}%
	\begin{tablenotes}    
		\footnotesize               
		\item[1] The values shown in bold are the optimal model indexes (except OLS algorithm) in each case.        
	\end{tablenotes}            
\end{threeparttable} 
	\label{e3-b500}%
\end{table}%

\begin{table}[htbp]
	\centering
	\caption{Average Running Time in Experiment 3 for $b=1000$}
	\begin{threeparttable}
				\begin{tabular}{c|cccccc}
				\hline
				\multicolumn{7}{c}{$b=1000$} \\
				\hline
				& case 1 & case 2 & case 3 & case 4 & case 5 & case 6 \\
				\hline
				& \multicolumn{6}{c}{homogeneous} \\
				\hline
				OLS   & 1.00480  & 0.99670  & 0.99650  & 0.99950  & 1.00350  & 1.00390  \\
				RLS   & 6.69460  & 6.77830  & 6.39620  & 4.47510  & 4.64710  & 4.46410  \\
				OHR   & \textbf{2.30300 } & \textbf{2.60550 } & \textbf{2.77850 } & \textbf{2.53540 } & \textbf{2.62030 } & \textbf{2.61390 } \\
				DC-HR & 2.73700  & 2.88140  & 2.83060  & 2.94160  & 2.70820  & 2.89440  \\
				UHR   & 3.06820  & 3.20160  & 3.01010  & 2.72710  & 2.86340  & 2.83450  \\
				\hline
				& \multicolumn{6}{c}{heterogeneous} \\
				\hline
				OLS   & 1.02390  & 0.98440  & 1.09730  & 1.01810  & 1.01980  & 1.01280  \\
				RLS   & 3.54000  & 3.19630  & 3.15630  & 3.24660  & 3.16610  & 3.26170  \\
				OHR   & \textbf{2.47570 } & \textbf{2.59900 } & \textbf{2.36000 } & \textbf{2.62570 } & 4.64180  & \textbf{2.31450 } \\
				DC-HR & 2.71570  & 2.92090  & 2.62800  & 2.97190  & \textbf{3.29770 } & 2.67450  \\
				UHR   & 2.50870  & 2.66040  & 2.35530  & 2.59340  & 2.87450  & 2.27890  \\
				\hline
			\end{tabular}%
	\begin{tablenotes}    
		\footnotesize               
		\item[1] The values shown in bold are the optimal model indexes (except OLS algorithm) in each case.        
	\end{tablenotes}            
	\end{threeparttable} 
	\label{e3-b1000}%
\end{table}%

\begin{figure}
	\centerline{\includegraphics[height=6.5cm]{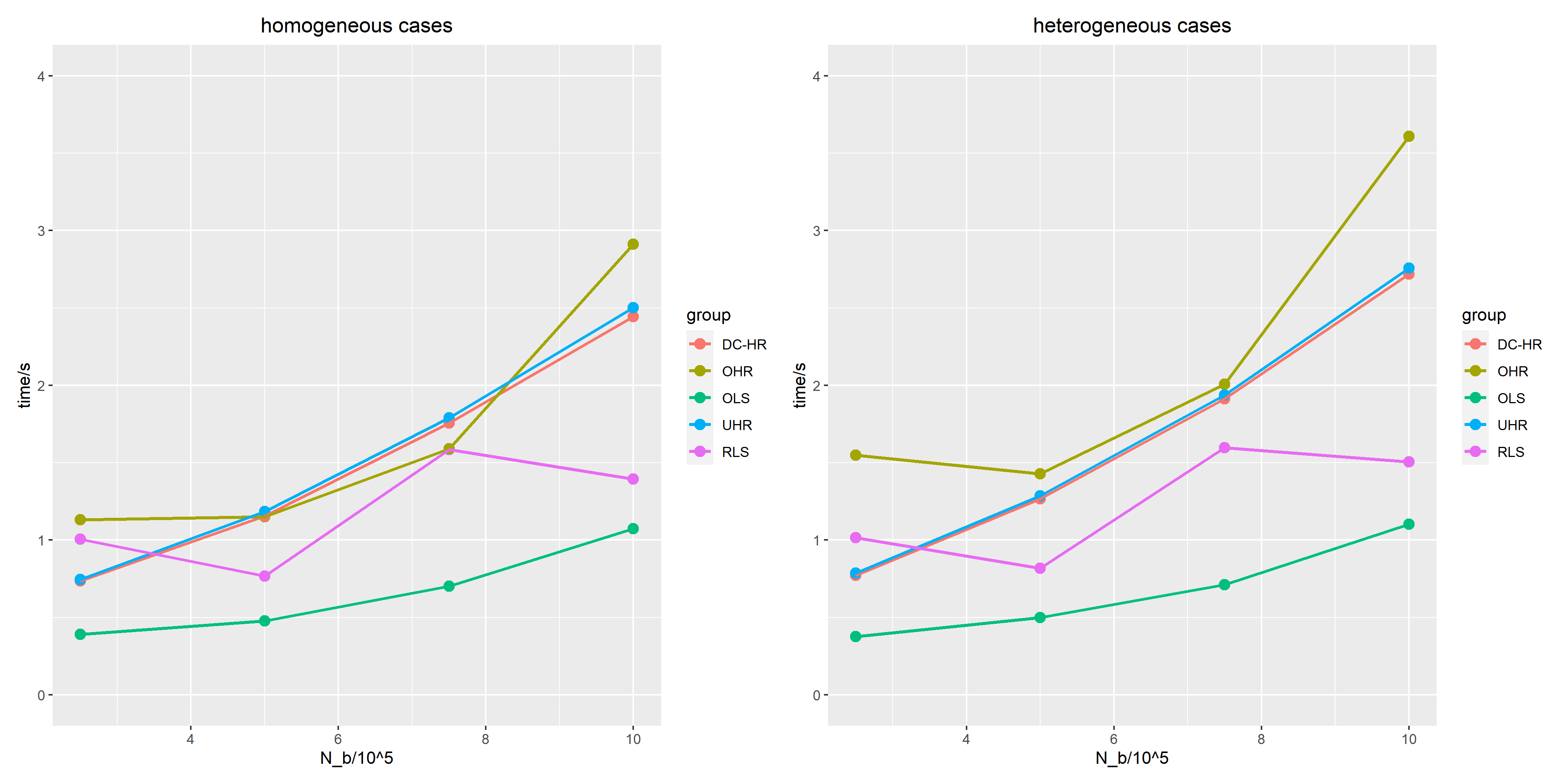}}
	\caption{total sample size-calculation time with $n_t=5000$}
	\label{graph3}
\end{figure}

\begin{figure}
	\centerline{\includegraphics[height=6.5cm]{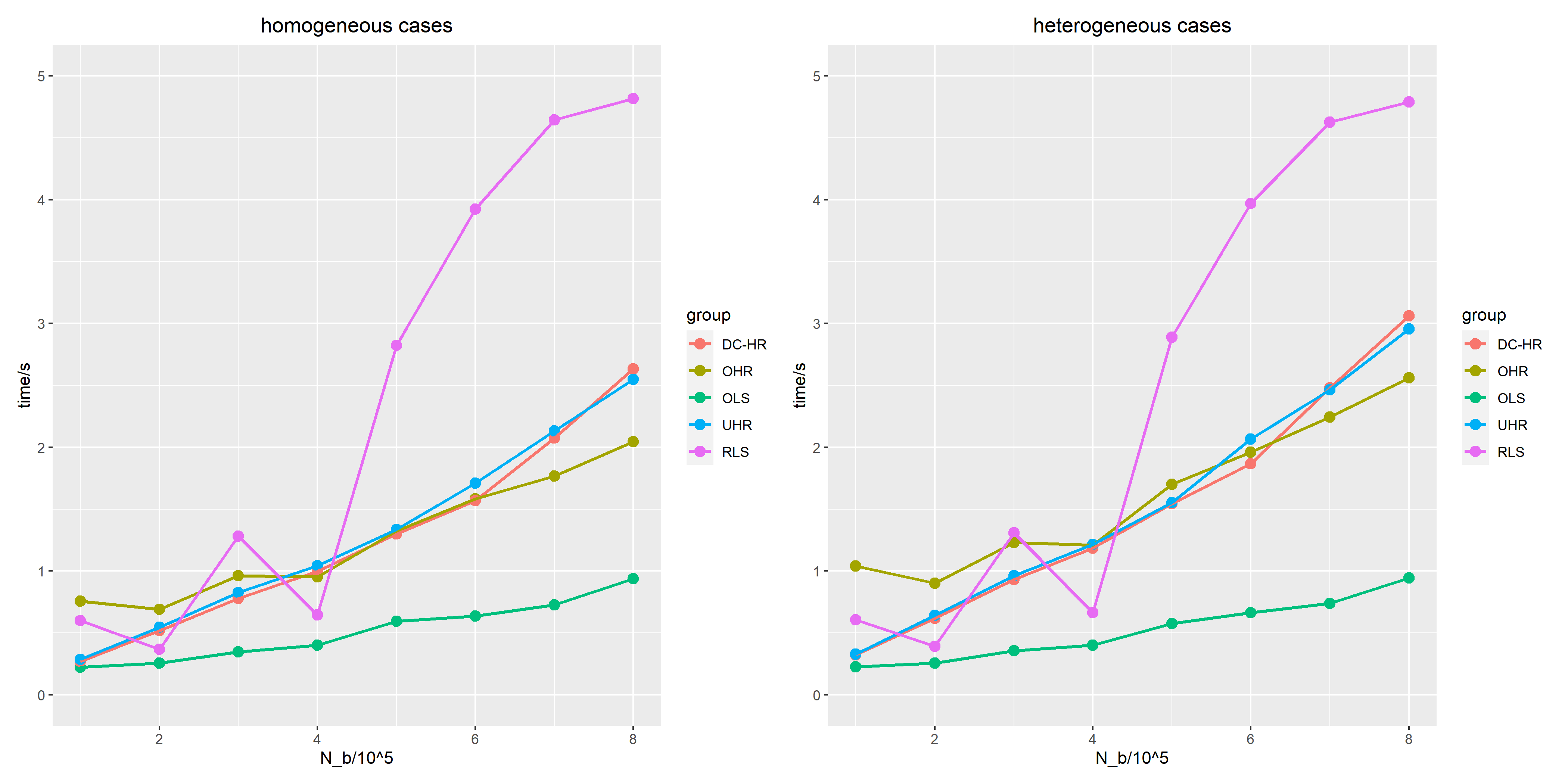}}
	\caption{total sample size-calculation time with $n_t=1000$}
	\label{graph4}
\end{figure}

For \textbf{Experiment\ 3} and \textbf{Experiment\ 4}, we divide them into homogeneous and heterogeneous cases and average running time of 6 cases of random error. The relationship between sample size and running time are represented in Figure \ref{graph3} and Figure \ref{graph4}. According to Figure \ref{graph4}, We can draw conclusions as follows.

(1) Overall, Linear Regression algorithms run faster than Huber algorithms, indicating the same result with \textbf{Experiment\ 2}. Nevertheless, estimation accuracy is the priority, so that we still prefer Huber algorithms, although Linear Regression algorithms have better communication efficiency.

(2) For Huber algorithms, UHR performs better than OHR in most cases in calculation efficiency. According to the line trend, running time of OHR climb faster than UHR when total sample size $N_b$ increasing. Moreover, we obtained the computation complexity of UHR and OHR before which are $O(KpNn)$ and $O(KpN^2)$. This is coordinated with the line in Figure \ref{graph3} that line of UHR and OHR are rather similar to linear and quadratic functions of $N_b$, which verifies UHR has lower computation efficiency than OHR in simulations.

Unlike $n_t=5000$ in \textbf{Experiment\ 3}, \textbf{Experiment\ 4} sets $n_t=1000$. Therefore, Figure \ref{graph4} shows a distinct graph from Figure \ref{graph3}. When $N_b<4\times10^5$, UHR in Figure \ref{graph4} still runs faster than OHR, presenting the same outcome with Figure \ref{graph3}. However, UHR shows a disadvantage in calculation gradually when $N_b>4\times10^5$. This result is not owing to the increase in $N_b$. For this deduction, we can refer to \textbf{Experiment\ 3}, where UHR runs faster than OHR with $N_b=1\times10^6,n_t=5000$. Therefore, we may assume bad performance in \textbf{Experiment\ 4} is due to another variant $b$. This is similar to observations in \textbf{Experiment\ 1}, which may result from increase in iterations. A possible solution is using parallel or distributed calculation on multiple machines inside the algorithm to cut down total running time.

\subsection{Real Data Analysis}

In real data analysis, we use our UHR, OLS, RLS, OHR, DC-HR algorithms used in simulation and DC-AHR algorithm to do the regression. Among them, DC-AHR is a combination of na\"ive-dc and Adaptive Huber regression\citep{sunAdaptiveHuberRegression2020,wangNewPrincipleTuningFree2021}. By comparing the results, we want to prove that our algorithm has good performance in real applications as well. 

Our real data is the airline dataset from 2009 ASA Data Expo(http://statcomputing.org/dataexpo/\\2009/the-data.html). It contains departure and landing information of 12 million business flights in America between 1978 and 2008. We choose data in 2007 as research object, which has 7,453,215 pieces of information, about 0.654GB in total. After removing the canceled and missing flight information, 7275288 samples were left. We use the first 3,000,000 samples as training set, and the rest as test set. Set subset number $b=100$, a rather small value, to better imitate the process of big data streams.

Model researched is a four element linear regression model proposed by \citeauthor{schifanoOnlineUpdatingStatistical2016}, which is widely used in many big data regression investigations\citep{jiangAdaptiveQuantileRegressions2021,wangStatisticalMethodsComputing2016, jiangCompositeQuantileRegression2018}.

\begin{equation}\label{real-model}
	AD=\gamma_0+\gamma_1 HD+\gamma_2 DIS+\gamma_3 NF+\gamma_4 WF+\epsilon
\end{equation} 

In Eq.(\ref{real-model}), $AD$ is a transformation of raw data variable $\rm{ArrDelay}$, standing for arrival delay. It is a continuous variable and transformed by formula ${\rm{log}(\bm{ArrDelay}-\rm{min}(\bm{ArrDelay})+1)}$. $HD$ stands for the hour of departure time, which is a discrete variable ranging from 1 to 24. $DIS$ is flight distance with unit of a thousand miles. $NF$ and $WF$ are both dummy variables. $NF$ stands for whether taking off at night. It equals to 1 for taking off during 8 pm to 5 am of the next day and equals to 0 else. $WF$ equals to 1 for taking off on weekends and equals to 0 on weekdays. 

We choose 3 evaluation indexes. The first is regression coefficient $\bm{\gamma}=(\gamma_0,\gamma_1,\gamma_2,\gamma_3,\gamma_4)^T$. By comparing itself and its significance, differences between 6 algorithms can be shown. It should be mentioned that in significance tests, we use bootstrap to estimate stand error and then calculate $t$-value, for in distributed algorithms, $t$-value cannot be obtained directly like Oralce ones. The second index is regression error including $mse$ and $mae$, which are both out of sample errors. Their expressions are like Eq.(\ref{real-mse}) and Eq.(\ref{real-mae}). The last index is running time, aiming at calculation efficiency.

\begin{equation}\label{real-mse}
	mse=\frac{1}{n}\sum_{i=1}^{n}(\widehat{y}_i-y_i)^2
\end{equation}

\begin{equation}\label{real-mae}
	mae=\frac{1}{n}\sum_{i=1}^{n}|\widehat{y}_i-y_i|
\end{equation}

The outcomes are presented in Table \ref{real-data}. Firstly, there is little difference between coefficients of each algorithm. The $t$-values are all significant, showing these methods are all interpretative to this model. In estimation efficiency, the estimation error of UHR is the smallest and also rather close to that of Oracle Huber, which again proves its asymptotic equivalency with Oracle algorithm. Meanwhile, its estimation error is smaller than that of DC-HR and DC-AHR as well, attesting Online Updating's better performance than divide-and-conquer. As for calculation efficiency, Huber algorithms run slower than Linear Regression algorithms. However, according to $mse$ and $mae$, Huber algorithms can cut down estimation error validly compared to Linear Regression. This is the reason we prefer Huber ones. Moreover, UHR has an outstanding performance between Huber ones with 5 seconds faster than Oracle Huber regression. This achieves our goal of speeding the calculation process. Overall, UHR performs impressively in real data analysis.

\begin{table}[htbp]
	\centering
	\caption{Results of Real Data Analysis}
	\begin{threeparttable}
	\begin{tabular}{c|c|c|c|c|c|c|c|c}
		\hline
		& (Intercept) & DepHour & Kdis  & night & weekend & $mse$   & $mae$   & \textit{time} \\
		\hline
		\multirow{2}[2]{*}{OLS} & 5.72482 & 0.00356 & -0.00194 & 0.02196 & -0.00754 & \multirow{2}[2]{*}{0.00905} & \multirow{2}[2]{*}{0.0882} & \multirow{2}[2]{*}{1.62} \\
		& (30937.344) & (283.338) & (-19.550)  & (121.83) & (-59.634)  &       &       &  \\
		\hline
		\multirow{2}[2]{*}{RLS} & 5.72482 & 0.00356 & -0.00194 & 0.02196 & -0.00754 & \multirow{2}[2]{*}{0.00905} & \multirow{2}[2]{*}{0.0882} & \multirow{2}[2]{*}{\textbf{0.695}} \\
		& (30455.939) & (247.18) & (-18.169)  & (89.745) & (-61.248)  &       &       &  \\
		\hline
		\multirow{2}[2]{*}{OHR} & 5.72547 & 0.00226 & -0.00147 & 0.00799 & -0.00549 & \multirow{2}[2]{*}{0.00832} & \multirow{2}[2]{*}{0.08129} & \multirow{2}[2]{*}{13.168} \\
		& (50804.848) & (295.477) & (-24.236)  & (72.776) & (-71.197)  &       &       &  \\
		\hline
		\multirow{2}[2]{*}{DC-HR} & 5.72599 & 0.00232 & -0.00131 & 0.00844 & -0.00557 & \multirow{2}[2]{*}{0.00831} & \multirow{2}[2]{*}{0.08113} & \multirow{2}[2]{*}{8.273} \\
		& (54921.000) & (301.501) & (-18.600)  & (61.269) & (-75.822)   &       &       &  \\
		\hline
		DC-AHR & 5.72582 & 0.00228 & -0.00118 & 0.00826 & -0.00549 & 0.00829 & 0.08096 & 134.377 \\
		\hline
		\multirow{2}[2]{*}{UHR} & 5.7256 & 0.00231 & -0.00146 & 0.00886 & -0.00546 & \multirow{2}[2]{*}{\textbf{0.00829}} & \multirow{2}[2]{*}{\textbf{0.08092}} & \multirow{2}[2]{*}{8.653} \\
		& (55172.597) & (289.168) & (-21.160)  & (67.636) & (-70.772)  &       &       &  \\
		\hline
	\end{tabular}%
	\begin{tablenotes}    
		\footnotesize               
		\item[1] The values shown in bold are the optimal model indexes in each case.        
		\item[2] The index in parentheses is the $t$-value.
		\item[3] The $t$-value of DC-AHR is not reported because it takes too long to calculate the $t$-value.
	\end{tablenotes}            
	\end{threeparttable} 
	\label{real-data}%
\end{table}%

\section{Conclusions and Discussions}

In order to process statistical analysis on big data streams and solve the outlier problem simultaneously, this paper proposes an Online Updating Huber Regression algorithm. By combining online updating method with Huber regression, UHR achieves both big data streams regression through updating historical data continuously and robust estimation on contaminated datasets. Proved by theoretical and simulation results, UHR is asymptotic equivalent to Oracle one using entire dataset and has lower computation complexity. After constructing our algorithm, we also apply it to simulations and real data analysis. In simulations, UHR performs outstandingly when random error is heavy-tailed distributed or has a large amount of outliers. It has good performance in calculation efficiency compared with the Oralce one as well, especially for cases when subset number $b$ is relatively small. In real data analysis, UHR also has good regression results for the airline dataset with the smallest estimation error, much faster calculation speed than the Oracle algorithm and significant regression coefficients, attesting its feasibility in real applications.

However, there are also some issues needed to be further investigated in the future. Firstly, we assume the true value of unknown parameters to be estimated does not change in data generating process. This assumption ignores to consider concept drift issue in big data streams, which may not be true in real applications. Thus, more complicated streaming data models need to be investigated. Secondly, in the concept of big data, we only consider big data with high volume but not with high dimensions. In future work, by combining it with the penalty function, we can investigate the high-dimensional regularized online updating problems. Lastly, the proposed algorithm is designed for independent data, which is a rather simple assumption. We can further focus on solutions to dependent or non-stationary big data streams in the future.

\bibliography{reference}  

\begin{thebibliography}{37}
\providecommand{\natexlab}[1]{#1}
\providecommand{\url}[1]{\texttt{#1}}
\expandafter\ifx\csname urlstyle\endcsname\relax
  \providecommand{\doi}[1]{doi: #1}\else
  \providecommand{\doi}{doi: \begingroup \urlstyle{rm}\Url}\fi

\bibitem[Allison(1995)]{1995The}
P.~D. Allison.
\newblock The impact of random predictors on comparisons of coefficients
  between models: Comment on clogg, petkova, and haritou.
\newblock \emph{American Journal of Sociology}, 100\penalty0 (5):\penalty0
  1294--1305, 1995.

\bibitem[Bassett and Jr.(1978)]{1978Regression}
Koenker~Gilbert Bassett and Jr.
\newblock Regression quantiles.
\newblock \emph{Econometrica}, 46\penalty0 (1):\penalty0 33--50, 1978.

\bibitem[Battey et~al.(2018)Battey, Fan, Liu, Lu, and
  Zhu]{batteyDistributedTestingEstimation2018}
Heather Battey, Jianqing Fan, Han Liu, Junwei Lu, and Ziwei Zhu.
\newblock Distributed testing and estimation under sparse high dimensional
  models.
\newblock \emph{The Annals of Statistics}, 46\penalty0 (3), June 2018.
\newblock ISSN 0090-5364.
\newblock \doi{10.1214/17-AOS1587}.

\bibitem[Chen and Zhou(2020)]{chenQuantileRegressionBig2020}
Lanjue Chen and Yong Zhou.
\newblock Quantile regression in big data: {{A}} divide and conquer based
  strategy.
\newblock \emph{Computational Statistics \& Data Analysis}, 144:\penalty0
  106892, April 2020.
\newblock ISSN 01679473.
\newblock \doi{10.1016/j.csda.2019.106892}.

\bibitem[Chen et~al.(2019)Chen, Liu, and
  Zhang]{chenQuantileRegressionMemory2019}
Xi~Chen, Weidong Liu, and Yichen Zhang.
\newblock Quantile regression under memory constraint.
\newblock \emph{The Annals of Statistics}, 47\penalty0 (6), December 2019.
\newblock ISSN 0090-5364.
\newblock \doi{10.1214/18-AOS1777}.

\bibitem[Chen and Xie(2014)]{chenSplitandconquerApproachAnalysis2014}
Xueying Chen and Min-ge Xie.
\newblock A split-and-conquer approach for analysis of.
\newblock \emph{Statistica Sinica}, 2014.
\newblock ISSN 10170405.
\newblock \doi{10.5705/ss.2013.088}.

\bibitem[Clogg et~al.(1995)Clogg, Clifford, C., Petkova, Eva, Haritou, and
  Adamantios]{Clogg1995Statistical}
Clogg, Clifford, C., Petkova, Eva, Haritou, and Adamantios.
\newblock Statistical methods for comparing regression coefficients between
  models.
\newblock \emph{American Journal of Sociology}, 1995.

\bibitem[Faming et~al.(2013)Faming, Liang, Yichen, Cheng, Qifan, Song,
  Jincheol, Park, Ping, and Yang]{Faming2013A}
Faming, Liang, Yichen, Cheng, Qifan, Song, Jincheol, Park, Ping, and Yang.
\newblock A resampling-based stochastic approximation method for analysis of
  large geostatistical data.
\newblock \emph{Jasa Journal of the American Statistical Association}, 2013.

\bibitem[Fang~Yao(2021)]{fangyaoOnlineEstimationFunctional2021}
Ying~Yang Fang~Yao.
\newblock Online {{Estimation}} for {{Functional Data}}.
\newblock \emph{Journal of the American Statistical Association}, pages 1--15,
  November 2021.
\newblock ISSN 0162-1459, 1537-274X.
\newblock \doi{10.1080/01621459.2021.2002158}.

\bibitem[Hu et~al.(2021)Hu, Jiao, Liu, Shi, and
  Wu]{huDistributedQuantileRegression2021}
Aijun Hu, Yuling Jiao, Yanyan Liu, Yueyong Shi, and Yuanshan Wu.
\newblock Distributed quantile regression for massive heterogeneous data.
\newblock \emph{Neurocomputing}, 448:\penalty0 249--262, August 2021.
\newblock ISSN 09252312.
\newblock \doi{10.1016/j.neucom.2021.03.041}.

\bibitem[Huber and Peter(1964)]{Huber1964Robust}
Huber and J.~Peter.
\newblock Robust estimation of a location parameter.
\newblock \emph{The Annals of Mathematical Statistics}, 35\penalty0
  (1):\penalty0 73--101, 1964.

\bibitem[Huber and Peter(1981)]{Huber1981}
Huber and J.~Peter.
\newblock [wiley series in probability and statistics] robust statistics
  (huber/robust statistics) || references.
\newblock pages 294--300, 1981.

\bibitem[Jiang et~al.(2018)Jiang, Hu, Yu, and
  Qian]{jiangCompositeQuantileRegression2018}
Rong Jiang, Xueping Hu, Keming Yu, and Weimin Qian.
\newblock Composite quantile regression for massive datasets.
\newblock \emph{Statistics}, 52\penalty0 (5):\penalty0 980--1004, September
  2018.
\newblock ISSN 0233-1888, 1029-4910.
\newblock \doi{10.1080/02331888.2018.1500579}.

\bibitem[Jiang et~al.(2021)Jiang, Chen, and
  Liu]{jiangAdaptiveQuantileRegressions2021}
Rong Jiang, Wei-wei Chen, and Xin Liu.
\newblock Adaptive quantile regressions for massive datasets.
\newblock \emph{Statistical Papers}, 62\penalty0 (4):\penalty0 1981--1995,
  August 2021.
\newblock ISSN 0932-5026, 1613-9798.
\newblock \doi{10.1007/s00362-020-01170-8}.

\bibitem[Jin et~al.(2015)Jin, Wah, Cheng, and
  Wang]{jinSignificanceChallengesBig2015}
Xiaolong Jin, Benjamin~W. Wah, Xueqi Cheng, and Yuanzhuo Wang.
\newblock Significance and {{Challenges}} of {{Big Data Research}}.
\newblock \emph{Big Data Research}, 2\penalty0 (2):\penalty0 59--64, June 2015.
\newblock ISSN 22145796.
\newblock \doi{10.1016/j.bdr.2015.01.006}.

\bibitem[Jordan et~al.(2019)Jordan, Lee, and
  Yang]{jordanCommunicationEfficientDistributedStatistical2019}
Michael~I. Jordan, Jason~D. Lee, and Yun Yang.
\newblock Communication-{{Efficient Distributed Statistical Inference}}.
\newblock \emph{Journal of the American Statistical Association}, 114\penalty0
  (526):\penalty0 668--681, April 2019.
\newblock ISSN 0162-1459, 1537-274X.
\newblock \doi{10.1080/01621459.2018.1429274}.

\bibitem[Kleiner et~al.(2014)Kleiner, Talwalkar, Sarkar, and
  Jordan]{kleinerScalableBootstrapMassive2014}
Ariel Kleiner, Ameet Talwalkar, Purnamrita Sarkar, and Michael~I. Jordan.
\newblock A scalable bootstrap for massive data.
\newblock \emph{Journal of the Royal Statistical Society: Series B (Statistical
  Methodology)}, 76\penalty0 (4):\penalty0 795--816, September 2014.
\newblock ISSN 13697412.
\newblock \doi{10.1111/rssb.12050}.

\bibitem[Lee et~al.(2015)Lee, Sun, Liu, and Taylor]{2015Communication}
J.~D. Lee, Y.~Sun, Q.~Liu, and J.~E. Taylor.
\newblock Communication-efficient sparse regression: a one-shot approach.
\newblock \emph{Computer Science}, 2015.

\bibitem[Lee et~al.(2020)Lee, Wang, and Schifano]{leeOnlineUpdatingMethod2020}
JooChul Lee, HaiYing Wang, and Elizabeth~D. Schifano.
\newblock Online updating method to correct for measurement error in big data
  streams.
\newblock \emph{Computational Statistics \& Data Analysis}, 149:\penalty0
  106976, September 2020.
\newblock ISSN 01679473.
\newblock \doi{10.1016/j.csda.2020.106976}.

\bibitem[Lian and Fan(2018)]{2018Divide}
H.~Lian and Z.~Fan.
\newblock Divide-and-conquer for debiased l1-norm support vector machine in
  ultra-high dimensions.
\newblock \emph{Journal of Machine Learning Research}, 18:\penalty0 1--26,
  2018.

\bibitem[Liang et~al.(2016)Liang, Kim, and Song]{2016AB}
Faming Liang, Jinsu Kim, and Qifan Song.
\newblock A bootstrap metropolis–hastings algorithm for bayesian analysis of
  big data.
\newblock \emph{Technometrics A Journal of Statistics for the Physical Chemical
  \& Engineering Sciences}, page 604, 2016.

\bibitem[Luo et~al.(2022)Luo, Sun, and Zhou]{luoDistributedAdaptiveHuber2022}
Jiyu Luo, Qiang Sun, and Wen-Xin Zhou.
\newblock Distributed adaptive {{Huber}} regression.
\newblock \emph{Computational Statistics \& Data Analysis}, 169:\penalty0
  107419, May 2022.
\newblock ISSN 01679473.
\newblock \doi{10.1016/j.csda.2021.107419}.

\bibitem[Luo and Song(2020)]{luoRenewableEstimationIncremental2020a}
Lan Luo and Peter X.-K. Song.
\newblock Renewable estimation and incremental inference in generalized linear
  models with streaming data sets.
\newblock \emph{Journal of the Royal Statistical Society: Series B (Statistical
  Methodology)}, 82\penalty0 (1):\penalty0 69--97, February 2020.
\newblock ISSN 13697412.
\newblock \doi{10.1111/rssb.12352}.

\bibitem[Nan and Xi(2011)]{2011Aggregated}
L.~Nan and R.~Xi.
\newblock Aggregated estimating equation estimation.
\newblock \emph{Statistics and its interface}, 1\penalty0 (1):\penalty0 73--83,
  2011.

\bibitem[Ping and Sun(2015)]{2015Leveraging}
M.~Ping and X.~Sun.
\newblock Leveraging for big data regression.
\newblock \emph{Wiley Interdisciplinary Reviews: Computational Statistics},
  7\penalty0 (1), 2015.

\bibitem[Schifano et~al.(2016)Schifano, Wu, Wang, Yan, and
  Chen]{schifanoOnlineUpdatingStatistical2016}
Elizabeth~D. Schifano, Jing Wu, Chun Wang, Jun Yan, and Ming-Hui Chen.
\newblock Online {{Updating}} of {{Statistical Inference}} in the {{Big Data
  Setting}}.
\newblock \emph{Technometrics}, 58\penalty0 (3):\penalty0 393--403, July 2016.
\newblock ISSN 0040-1706, 1537-2723.
\newblock \doi{10.1080/00401706.2016.1142900}.

\bibitem[Sun et~al.(2020)Sun, Zhou, and Fan]{sunAdaptiveHuberRegression2020}
Qiang Sun, Wen-Xin Zhou, and Jianqing Fan.
\newblock Adaptive {{Huber Regression}}.
\newblock \emph{Journal of the American Statistical Association}, 115\penalty0
  (529):\penalty0 254--265, January 2020.
\newblock ISSN 0162-1459, 1537-274X.
\newblock \doi{10.1080/01621459.2018.1543124}.

\bibitem[Wang et~al.(2016)Wang, Chen, Schifano, Wu, and
  Yan]{wangStatisticalMethodsComputing2016}
Chun Wang, Ming-Hui Chen, Elizabeth Schifano, Jing Wu, and Jun Yan.
\newblock Statistical {{Methods}} and {{Computing}} for {{Big Data}}.
\newblock \emph{Statistics and Its Interface}, 9\penalty0 (4):\penalty0
  399--414, 2016.
\newblock ISSN 19387989, 19387997.
\newblock \doi{10.4310/SII.2016.v9.n4.a1}.

\bibitem[Wang et~al.(2018)Wang, Chen, Wu, Yan, Zhang, and
  Schifano]{wangOnlineUpdatingMethod2018}
Chun Wang, Ming-Hui Chen, Jing Wu, Jun Yan, Yuping Zhang, and Elizabeth
  Schifano.
\newblock Online updating method with new variables for big data streams.
\newblock \emph{Canadian Journal of Statistics}, 46\penalty0 (1):\penalty0
  123--146, March 2018.
\newblock ISSN 03195724.
\newblock \doi{10.1002/cjs.11330}.

\bibitem[Wang et~al.(2022)Wang, Wang, and
  Li]{wangRenewableQuantileRegression2022}
Kangning Wang, Hongwei Wang, and Shaomin Li.
\newblock Renewable quantile regression for streaming datasets.
\newblock \emph{Knowledge-Based Systems}, 235:\penalty0 107675, January 2022.
\newblock ISSN 09507051.
\newblock \doi{10.1016/j.knosys.2021.107675}.

\bibitem[Wang et~al.(2021)Wang, Zheng, Zhou, and
  Zhou]{wangNewPrincipleTuningFree2021}
Lili Wang, Chao Zheng, Wen Zhou, and Wen-Xin Zhou.
\newblock A {{New Principle}} for {{Tuning-Free Huber Regression}}.
\newblock \emph{Statistica Sinica}, 2021.
\newblock ISSN 10170405.
\newblock \doi{10.5705/ss.202019.0045}.

\bibitem[Wu et~al.(2021)Wu, Chen, Schifano, and
  Yan]{wuOnlineUpdatingSurvival2021}
Jing Wu, Ming-Hui Chen, Elizabeth~D. Schifano, and Jun Yan.
\newblock Online {{Updating}} of {{Survival Analysis}}.
\newblock \emph{Journal of Computational and Graphical Statistics}, 30\penalty0
  (4):\penalty0 1209--1223, October 2021.
\newblock ISSN 1061-8600, 1537-2715.
\newblock \doi{10.1080/10618600.2020.1870481}.

\bibitem[Xue and Hu(2021)]{xueOnlineUpdatingInformation2021}
Yishu Xue and Guanyu Hu.
\newblock Online updating of information based model selection in the big data
  setting.
\newblock \emph{Communications in Statistics - Simulation and Computation},
  50\penalty0 (11):\penalty0 3516--3529, November 2021.
\newblock ISSN 0361-0918, 1532-4141.
\newblock \doi{10.1080/03610918.2019.1626886}.

\bibitem[Zhang et~al.(2013)Zhang, Duchi, and Wainwright]{2013Divide}
Y.~Zhang, J.~C. Duchi, and M.~J. Wainwright.
\newblock Divide and conquer kernel ridge regression: A distributed algorithm
  with minimax optimal rates.
\newblock \emph{Journal of Machine Learning Research}, 30\penalty0
  (1):\penalty0 592--617, 2013.

\bibitem[Zhang et~al.(2012)Zhang, Duchi, and
  Wainwright]{zhangCommunicationefficientAlgorithmsStatistical2012}
Yuchen Zhang, John~C. Duchi, and Martin~J. Wainwright.
\newblock Communication-efficient algorithms for statistical optimization.
\newblock In \emph{2012 {{IEEE}} 51st {{IEEE Conference}} on {{Decision}} and
  {{Control}} ({{CDC}})}, pages 6792--6792, {Maui, HI, USA}, December 2012.
  {IEEE}.
\newblock ISBN 978-1-4673-2066-5 978-1-4673-2065-8 978-1-4673-2063-4
  978-1-4673-2064-1.
\newblock \doi{10.1109/CDC.2012.6426691}.

\bibitem[Zhao et~al.(2016)Zhao, Cheng, and Liu]{2016A}
T.~Zhao, G.~Cheng, and H.~Liu.
\newblock A partially linear framework for massive heterogeneous data.
\newblock \emph{Annals of Statistics}, 44\penalty0 (4):\penalty0 1400--1437,
  2016.

\bibitem[Zhu et~al.(2021)Zhu, Li, and
  Wang]{zhuLeastSquareApproximationDistributed2021}
Xuening Zhu, Feng Li, and Hansheng Wang.
\newblock Least-{{Square Approximation}} for a {{Distributed System}}.
\newblock \emph{Journal of Computational and Graphical Statistics}, 30\penalty0
  (4):\penalty0 1004--1018, October 2021.
\newblock ISSN 1061-8600, 1537-2715.
\newblock \doi{10.1080/10618600.2021.1923517}.

\end{thebibliography}
\bibliographystyle{plainnat}

\end{document}